\documentclass[graybox]{svmult}

\usepackage{mathptmx}       
\usepackage{helvet}         
\usepackage{courier}        
\usepackage{type1cm}        
\usepackage{makeidx}         
\usepackage{graphicx}        
\usepackage{multicol}        
\usepackage[bottom]{footmisc}

\usepackage{cancel}

\newcommand{\oversim}[2]{\protect{\mbox{\lower0.5ex\vbox{%
   \baselineskip=0pt\lineskip=0.2ex
   \ialign{$\mathsurround=0pt #1\hfil##\hfil$\crcr#2\crcr\sim\crcr}}}}} 
\newcommand{\simgreat}{\mbox{$\,\mathrel{\mathpalette\oversim>}\,$}} 
\newcommand{\simless} {\mbox{$\,\mathrel{\mathpalette\oversim<}\,$}} 
\makeindex             

\begin{document}

\title*{A modern view of galaxies and their stellar populations}
\author{Pavel Kroupa\\
  {\small This text (minus the Abstract and the Conclusions) was
    published in 2016 in two chapters in the book ``{\it From the
      Realm of the Nebulae to Populations of Galaxies - Dialogues from
      a Century of Research}'', edited by Mauro D'Onofrio, Roberto
    Rampazzo \& Simone Zaggia, Springer International Publishing
    Switzerland, ISBN 978-3-319-31004-6, DOI
    10.1007/978-3-319-31006-0.  Selected by CHOICE magazine
in the list of the most ``Outstanding Academic Books'' of 2017.}}
\institute{Pavel Kroupa\\ Helmholtz-Institut f\"ur Strahlen- und
  Kernphysik, Universit\"at Bonn, Nussallee 14--16, D-53121 Bonn,
  Germany,\\Charles University,
Astronomicky ustav UK,
V Hole\v{s}ovi\v{c}k\'ach 2,
CZ-18000 Praha,
Czech Republic\\
  \email{pkroupa@uni-bonn.de} }
%
%
\maketitle

\abstract{A critical discourse is provided on the current status of
  the astrophysics of galaxies in view of open fundamental questions
  on the law of gravitation and the physics-driven variation of the
  galaxy-wide stellar initial mass function (GWIMF).  The
  Einstein/Newtonian plus cold or warm dark-matter-based models face
  many significant but unresolved tensions (e.g. planes of satellites,
  the highly organised and symmetrical structure of the Local Group,
  the local Gpc-scale void).  The accumulating nature of these
  indicates quite compellingly the need for a different theoretical
  framework. An example is the prediction, made in 1997, of the
  existence of satellite galaxies with near-exact properties to the
  Hercules dwarf spheroidal satellite galaxy discovered in 2007 {\it
    if} it is a tidal dwarf galaxy void of dark matter.  Such results
  indicate that once the evidence is accepted that dark matter
  particles have no role in galaxies with gravitation being
  effectively Milgromian, then the resulting theoretical understanding
  of galaxies becomes much simpler and highly predictive with
  remarkable successes. Along with these scientific advancements, the
  recent observational data solidify the existence of one of the most
  important relations in star-formation theory, namely the
  $m_{\rm max}-M_{\rm ecl}$ relation. The data rule out stochastic
  star formation and confirm that the IMF becomes increasingly
  top-heavy with increasing star-formation rate density. Galaxies and
  their embedded-star-cluster building blocks are therefore
  self-regulated dynamical systems which are computationally
  accessible.  Finding a theory for the formation of galaxies involves
  the development of a new cosmological model, which may differ
  substantially from the dark-matter based one.  Very significant new
  opportunities have thus emerged for inquisitive and daring
  researchers which may appear risky now but almost certainly 
  lead to major breakthroughs. }

\section{The dark matter (DM) problem}
\label{sec:DM_pk}

\subsection{DM Question 1} 

{\it You clearly speak in several of your recent works of a ``Dark Matter
Crisis''. May you summarize your point of view on this subject, with
particular reference to the consequences of your ideas for the
commonly accepted view of galaxies as DM dominated systems}? 

\subsubsection{Answer}

The recent publications meant here are \cite{Kroupa10, Kroupa12a,
  Kroupa12b, Kroupa14}.  The argument is straight forward: In any
realistic cosmological model there are two types of dwarf galaxies
(the dual dwarf galaxy theorem). In the standard model of cosmology
(SMoC), the dual dwarf galaxy theorem implies a bifurcation: the one
type of (primordial) dwarf galaxy (PDG) resides in and is dominated by
a dark matter halo, while the other type of dwarf galaxy (tidal dwarf
galaxy, TDG) contains no significant amount of dark matter by the
nature of their formation in tidal tails drawn out in galaxy--galaxy
encounters. The dual dwarf galaxy theorem is falsified by the real
universe in that no evidence published to date can be found to
distinguish the dark matter content of these two different types of
dwarf galaxies. The explicit tests made are on the baryonic
Tully-Fisher relation and the radii of pressure supported dwarf
galaxies. Therewith, given the presently available data, the SMoC is
ruled out as a model of the real Universe:

\begin{equation}
{\rm SMoC}  \;  \Rightarrow \; {\rm PDGs} \ne {\rm TDGs}
\quad \Longleftrightarrow \quad
 {\rm PDGs} = {\rm TDGs} \; \Rightarrow \; \cancel  {\rm SMoC}.
\label{eq_pk:fals}
\end{equation}

\subsubsection{ Is this deduction correct?} 

Clearly, by deducing the SMoC to be invalid, I am arguing against the
main stream scientist, where the majority of colleagues appear to be
convinced that dark matter particles exist such that the SMoC is a
physically relevant model of the Universe. The conviction that this is
true is so powerful, that very major professorial chairs,
directorships and vast amounts of money are being expended to fund a
large number of research groups to work in the framework of this model
and to search for the dark matter particle(s).\footnote{Cases in point
  are found in footnote~8 in \cite{Kroupa14}:  \cite{Lang14} writes ``Due
  to a large number of astrophysical observations ... we know today
  that dark matter exists'' (originally: ``Aufgrund einer Vielzahl von
  astrophysikalischen Beobachtungen ... wissen wir heute, dass Dunkle
  Materie existiert'' ) and ``The question is thus not: does dark
  matter exist?  Rather, the issue is to find out: what does it
  consist of?''  (originally: ``Die Frage ist also l\"angst nicht
  mehr: Existiert die Dunkle Materie?  Vielmehr gilt es
  herauszufinden: Woraus besteht sie?''). A similar statement is found
  in chapter~25 of the Review of Particle Physics \cite{Olive14}:
  ``The existence of Dark (i.e., non-luminous and non-absorbing)
  Matter (DM) is by now well established'', although the correct
  statement should have been something like ``The existence of Dark
  (i.e., non-luminous and non-absorbing) Matter (DM) is at present a
  favored hypothesis''.}  With a deduction such as the above
(Eq.~\ref{eq_pk:fals}) I am putting my reputation at stake, so I need
to be completely convinced, without a shadow of a doubt, that my
conclusion is true: {\it If the deduction (Eq.~\ref{eq_pk:fals}) is
  wrong, then there should be indications in the astronomical data
  which would imply it to be wrong}. Is this the case? Or, do
independent data rather support the deduction?

In performing the consistency checking, it is useful to first
reconsider the foundations of the SMoC. The SMoC is based on assuming
that Albert Einstein's Theory of General Relativity (GR)
\cite{Einstein16} is valid everywhere. As discussed in the above
mentioned papers of mine, this assumption is found to quickly fail
multiply times leading to the discovery of new physics we refer to as
inflation, dark matter and dark energy (Eq.~\ref{eq_pk:theSMoC}
below). But is GR really correct?  I had no reason to doubt this
fundamental assumption, especially given the fundamental axioms such
as the equivalence principle it rests upon and the very accurate and
precise tests of it in the strong field regime (Solar System and
stronger in terms of acceleration) \cite{Byrd14}.  But in~2010 I
realized that Einstein had to work entirely with empirical constraints
on gravitation as available within the Solar System only, and that
tests performed many decades later of GR in the weak-field regime are
not conclusive, because the requirement to introduce dark matter
particles may merely be fixing a failure of GR.  That is, Einstein had
no other data at hand, apart from the precession of Mercury's orbit,
than those already used in the 17th century by Isaac Newton who
discovered his law of universal gravitation from such empirical
data. Galaxies had been realized as to what they are only decades
after GR was published, and the motion of matter within them was
mapped even later, at the end of the 1970s and early 1980s by Vera
Rubin and Albert Bosma. {\it Thus, assuming GR to be valid everywhere
  constitutes an extrapolation of an empirical law by many orders of
  magnitude beyond the experimental data range it was originally
  derived from}. This realisation led to a loss of confidence on my
behalve in the SMoC, making it all the more plausible, in my eyes,
that the above deduced falsification of the SMoC is correct, and that
the ``new physics'' mentioned above merely constitute mathematical
additions to reduce the tension between an extrapolation and the data.
The SMoC is discussed critically from a mathematical point of view by
e.g. \cite{KS14}.

\subsubsection {Consistency of the above deduction Eq.~\ref{eq_pk:fals}} 

The consistency of the above deduction is tested using independent
data in \cite{Kroupa10, Kroupa12a, Kroupa12b, Kroupa14}. Here a few
of the arguments are listed again:
\begin{itemize}

\item The arrangement of dwarf satellite galaxies in a rotating
  disk-of-satellites (DoS) around Andromeda or in a
  vast-polar-structures (VPOS) in the case of the Milky Way (MW) and
  other host galaxies within distances of 100 to 250~kpc rule out
  their origin as dark-matter sub-structures \cite{Pawlowski12b,
    Pawlowski14, Pawlowski14b, Ibata14, Ibata14b, Ibata14c}. But the
  dwarf satellite galaxies within and outside of the DoSs have
  indistinguishable physical properties \cite{Kroupa14, Collins14}
  such that their apparent dark matter content cannot be due to the
  presence of dark matter.

\item The lack of solutions with {\it Chandrasekhar dynamical friction} for
  dark-matter dominated dwarf galaxies orbiting in the putative dark
  matter halo of the Milky Way \cite{Angus11}. 

\item The lack of evidence for the merger activity \cite{Shankar14} in
  the galaxy population in that the vast number of galaxies brighter
  than about $10^{10}\,L_\odot$ \cite{Delgado10} are thin disk
  galaxies with radii extending to dozens of~kpc and the majority of
  these not even having a classical bulge \cite{Kormendy10}.

\item Related to this is the lack of variation of galaxies (they are
  all too similar as pointed out by Mike Disney \cite{Disney08}).

\item And, still concerning disk galaxies: the observed rotation
  curves are not matched by even the most advanced and the most recent
  hand-picked and fine-tuned modelling attempts within the SMoC
  framework \cite{WK15}.

\item The downsizing problem.

\item The Bullet cluster.

\item The local highly significant matter under-density on scales of
  10--300~Mpc (Fig.~\ref{fig_pk:underdensity}). 
\begin{figure*}
  \resizebox{11cm}{!}{
    \includegraphics{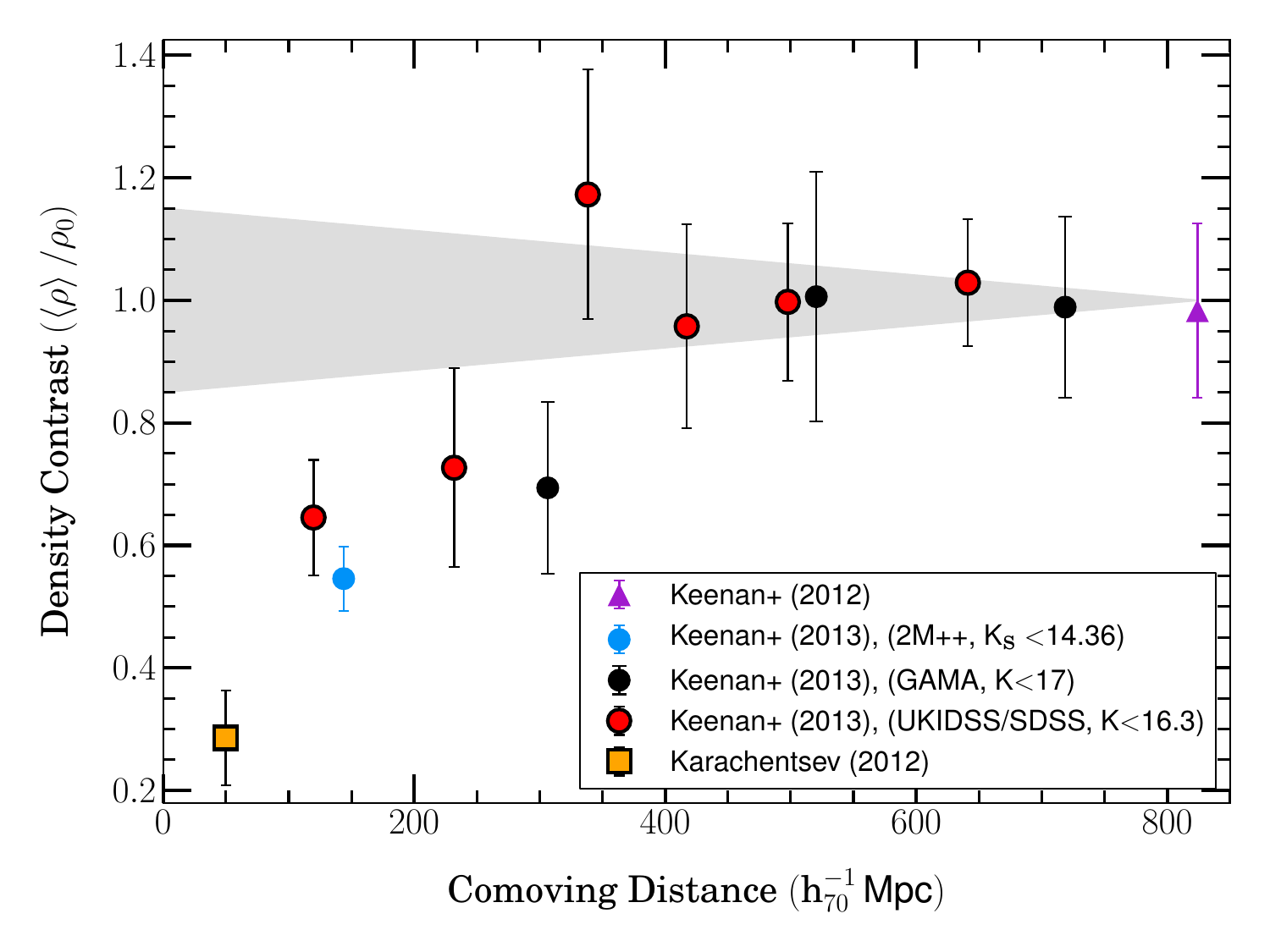}
  }
  \caption{ The observed significant underdensity of matter within
    about 300 Mpc of the Milky Way appears to be in extreme conflict
    with the SMoC which requires a value of one in the density
    contrast with fluctuations within the shaded region (for details
    see fig.1 in \cite{Kroupa14}).  }
  \label{fig_pk:underdensity}
\end{figure*}

\item And many more.

\item Not discussed in the above papers of mine is the additional evidence
  against extended heavy dark matter halos by the longness of tidal
  ails drawn out in galaxy--galaxy encounters \cite{DMH96, DMH99}.
  This is failure~23 (see the list of 22~failures in \cite{Kroupa12a}).

\item A particularly interesting and directly relevant very recent
  observational result, not mentioned in the above papers of mine, is
  that \cite{Lena14} have found that super-massive black holes
  (SMBHs) do not show the displacements they ought to show if merging
  were as common as is expected in the SMoC. Remember, in the absence
  of dark matter galaxies would not merge when they interact unless
  they have small, penetrating impact parameters \cite{Toomre77}.   This is failure~24.

\item An additional new failure of the SMoC is the finding that within
the local~10~Mpc volume there is a highly significant deficit of
massive dwarf galaxies \cite{Klypin14}.   This is failure~25.

\item A further additional new failure of the SMoC is the
  too-massive-to-fail problem for dwarf galaxies in voids
  \cite{Papastergis14}.   This is failure~26.

\item The highly symmetrical structure of the Local Group of galaxies
  (fig.~9 in Pawlowski et al. (2013, \cite{Pawlowski13b}) was also not
  known until~2013 and is well beyond being understood in any current
  theoretical framework, and is certainly entirely inconsistent with
  the Local Group being the result of a long history of mergers of
  dark matter halos.  This is failure~27. The related extraordinarily
  symmetrical structure of the Local Group is shown in
  Fig.~\ref{fig_pk:LG}.
  \begin{figure*}
  \centerline{
  \resizebox{15cm}{!}{
    \includegraphics{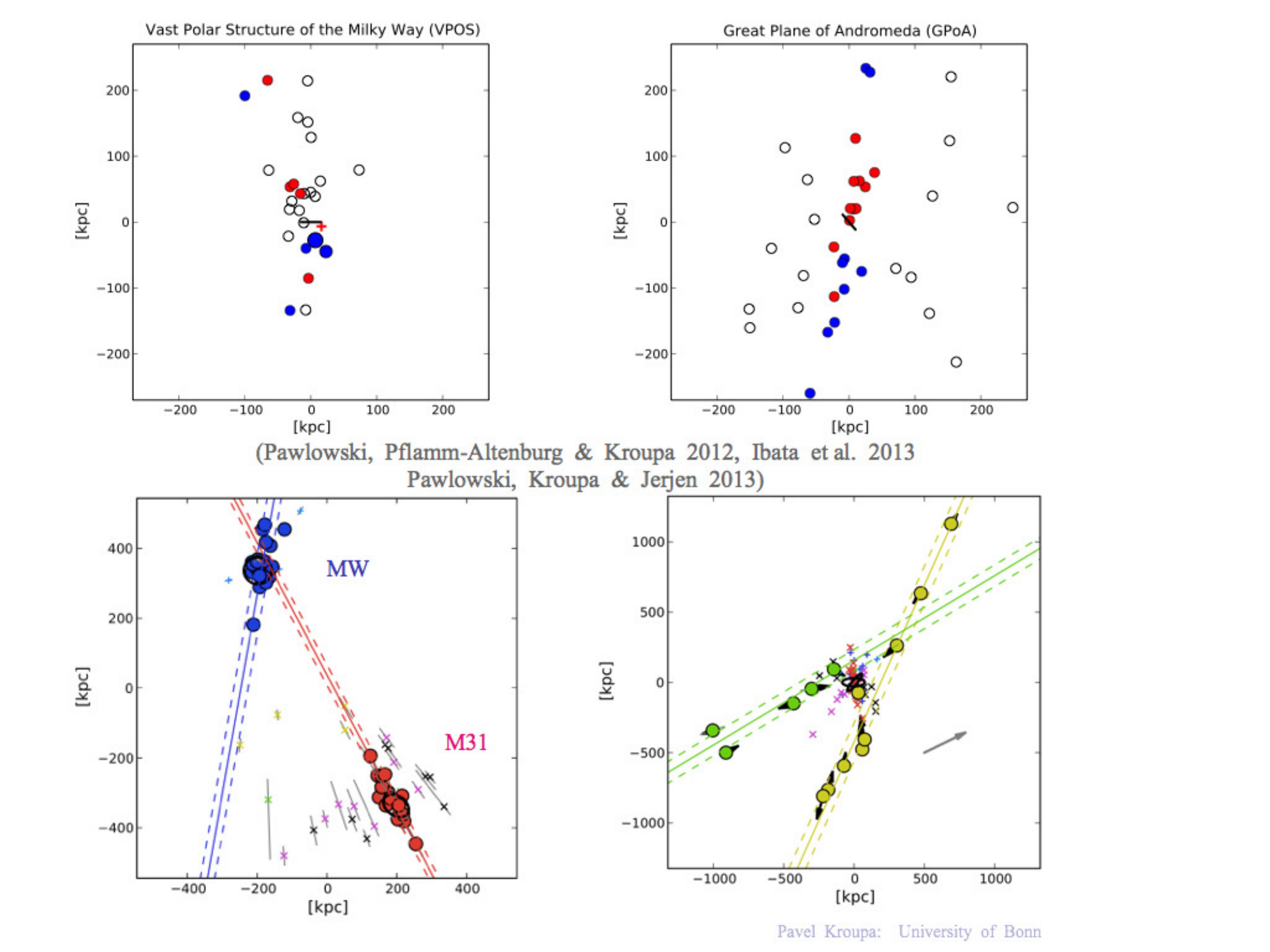}
  }}
  \caption{The Local group is highly organised: the satellite
    galaxies of the Milky Way are in a vast polar disk-like structure
    (blue ones are moving one way, red ones the other). Andromeda also
    has a great plane of satellites (blue satellites are moving in the
    same direction as the blue ones in the case of the Milky Way). The
    two satellite systems of the Milky Way and of Andromeda are
    aligned (lower left panel). All non-satellite galaxies in the
    Local Group are in two symmetric planes with near-equal properties
    (extension, thickness).  For details see fig.2 in
    \cite{Kroupa14b}.  }
  \label{fig_pk:LG}
\end{figure*}

\end{itemize}

None of these problems have ever been solved convincingly.
All tests performed are well consistent with the above conclusion
(Eq.~\ref{eq_pk:fals}), such that dynamically relevant particle dark
matter on galaxy scales cannot be present.  That evidence for
dynamical friction on the dark matter halos is absent nearly entirely
in the data solidifies this conclusion, which is also consistent with
the problem of observed tidal tails in interacting galaxies being much
longer than what they ought to be if the extensive and massive dark
mater halos were present.  To my astonishment, virtually every {\it
  prediction} made within the framework of the SMoC seems to have been
falsified by observation. This is visualized using the {\it Theory
  Confidence Graph} of \cite{Kroupa12a} (which does not include a
number of failures, such as the length-of-the tidal tails, the absence
of evidence for dynamical friction, the lack of recoiling black
holes).  While I am fully aware of many of my colleagues portraying
the converse of this, also in lectures to undergraduates, the
blatantly obvious disagreement, time and again, between the SMoC
results and predictions and the observational data I am finding
stunning.

If each ``problem'' (more correctly: failed prediction) is associated
with a loss of confidence of 50~(or 30)~per cent that the SMoC is
valid, then the overall remaining confidence becomes
$0.5^{22+5}=1/10^{8.13}$ (or $0.7^{27}=10^{-4.18}$), i.e. $7.5\times 10^{-7}$
(or $6.6\times 10^{-3}$)~per cent.
\begin{svgraybox}
  It follows that independently of the falsification of the SMoC
  through the dual dwarf galaxy theorem, all the 27 tests above lead
  to the SMoC being ruled out with at least 4~sigma, taking a
  conservative loss of confidence per failed prediction of only 30~per
  cent.
\end{svgraybox}

\subsubsection {Reactions} 

Historically, an interesting sociological process began acting in two
phases in repressing the notion that the SMoC may not be a valid
model. The primary phase is understandable conceptually because GR has
been found to be an excellent description of gravitational phenomena
where precise tests have been possible in the strong-field regime. Big
Bang cosmology based on GR is a convincing theory, and alternatives,
such as the steady-state model of the Universe, have been ruled out by
observation.

\begin{itemize}

\item The {\it primary phase} involves the additions of major new
  physics notions: The first failures emerged quickly but were
  considered to be evidence for new physics, which is a relevant
  scientific procedure.  Thus, in 1980 and 1998, respectively,
  inflation and dark energy (DE) were introduced as similar
  mathematical extensions of GR.  They are well motivated and solve a
  number of otherwise not understood observational problems.  In 1981
  the non- or at most weakly-interacting non-relativistic dark matter
  particles (DMPs) were introduced into the model in order to allow
  structures to grow in the diluted inflated model and to account for
  the observationally discovered mismatch between data and Newtonian
  gravitation on scales of galaxies and larger.  Thus the current
  mathematical formulation of the SMoC can be summarized as
\begin{equation}
{\rm SMoC} = {\rm GR} + {\rm inflation} + {\rm DE} + {\rm DMPs}.
\label{eq_pk:theSMoC}
\end{equation}
The philosophically unattractive aspects of these three major
extensions of the original model based on GR is that neither inflation
nor dark energy are understood theoretically thereby also displaying
important conceptual problems \cite{Baryshev06, Brandenberger08,
  Steinhardt09}. And, although the standard model of particle physics
(SMoPP) has been found to be an excellent description of nature apart
from gravity, dark matter particles are neither covered by it nor is
there direct experimental evidence for these particles despite a few
decade long and financially rather impressive effort.

\item The {\it secondary phase} began as further failures were
  becoming evident after the late 1980s, in that the observed galaxy
  population was not emerging in the SMoC. Astronomers began to argue
  that the physics of the known baryonic matter is so uncertain such
  that the structures driven by the supposedly well known and dominant
  but undetected dark matter cannot be quantified properly. Thus, for
  example, that the computed galaxy population consisted of largely
  barely rotating spheroidal or compact rotational systems in
  disagreement with the more than 90~per cent of all galaxies brighter
  than about $L>10^{10}\,L_\odot$ being large thin disk galaxies
  \cite{Delgado10} is being argued away by the baryons exerting an
  essentially unknown effect onto the dark matter particles such that
  they readjust in order to allow thin extended disk galaxies to
  grow. In truth it has only been possible to obtain models which
  roughly resemble real disk galaxies by hand-selecting individual
  dark matter halos without significant recent mergers \cite{WK15}.
  That the galaxy population calculated in the SMoC has severe
  problems in representing the real galaxy population is evident in
  the most recent very major and also celebrated computational
  Illustris Project\footnote{http://www.illustris-project.org/}, in
  which the feedback physics is modelled with the stellar wind speed
  being a few times the one-dimensional velocity dispersion of the
  dark matter halo within which they occur.  This is unphysical (see
  also \cite{Schaye15}, who apply a physically much more convincing
  feedback model), and such models are ruled out by the observation
  that TDGs do form and survive for Gyr \cite{Duc14} despite them not
  having dark matter halos \cite{BH92, Bournaud10}. In the end, which
  physical principle that may be derivable from the SMoPP or from dark
  matter physics can arrange such a coupling between baryons and dark
  matter which do not couple apart through gravitation and perhaps
  weakly? And, this would imply that present work on stellar winds and
  supernova explosions would be wrong since their hosting dark matter
  halos are not included in stellar astrophysical modelling.

  As another example, the {\it satellite galaxy over-prediction
    failure} became the {\it missing satellite problem} which is now
  deemed to be solved with such a wide range of post-dictions that any
  observation would be consistent with the model. According to the
  models by \cite{HWP14}, essentially any number of satellite galaxies
  will be consistent with the observed numbers, whatever they turn out
  to be.  Any baryonic-physics processes introduced and fine-tuned to
  solve some particular problem is thereby typically not checked for
  its affect on other issues. For example, ``solving'' the satellite
  over-prediction problem may lead to incompatibilities with the
  properties of disk galaxies.

  Related to this last issue and to how dark-matter computational
  astrophysics is applied to satellite galaxies and as an example of
  state-of-the art research, \cite{BB14} address the problem of
  whether ultra-compact dwarf galaxies (UCDs, i.e. essentially
  extremely massive globular clusters) may be the stripped nuclei of
  PDGs falling towards the centre of a galaxy cluster. They model the
  dark matter halo of the host, but neglect the dark matter halos of
  the PDGs, finding that they can be destroyed leaving their nuclei as
  posible UCD satellites. This work is now being cited as evidence
  that UCDs may be stripped nuclei in the framework of the SMoC. But
  this is wrong because the result is obtained using a highly
  inconsistent method: (i) if the SMoC is assumed to be valid (this is
  perfectly in order as an assumption, although being unrealistic)
  then the dwarfs would have large and massive dark matter halos
  \cite{Kroupa14} such that dynamical friction would have them merge
  with the host within a few orbital times and stripping of the nuclei
  would become unlikely because first the dwarf galaxy halo needs to
  be removed before it's stellar body is stripped. The authors have
  not shown this to be the case. (ii) If, on the other hand, the dark
  matter halo is neglected as the authors do (this is also an
  assumption which is admissible but with constraints) then the only
  physically relevant option the authors have is to switch to
  effectively non-Newtonian dynamics, because the rotation curves of
  galaxies need to be accounted for but without dark matter. But their
  integrations of the equations of motion are Newtonian. Thus, here
  too the authors have not shown that stripping works, since they
  computed the wrong problem. (iii) The computation the authors made
  and their assumption of neglecting the satellite galaxy dark matter
  halo might be argued to be correct within the SMoC if the satellite
  galaxy with its nucleus is a TDG, since TDGs do not have dark matter
  halos in the SMoC. However, this is a superficial view because this
  approach would also be fundamentally flawed by the dual dwarf galaxy
  theorem: if there are TDGs that play the role assumed in this
  problem, then their numbers would be so large (they'd be dE
  galaxies) that no room remains for the expected large numbers of
  PDGs. The points (i)--(iii) exemplify the type of internally
  inconcistent (but technically correct) research passing the
  peer-review process in this field of study. It is not the only case,
  and a reader might ask the question: {\it What has actually been shown
  by this type of work such that it can be applied to understanding
  the observations?}

\end{itemize}

Individual members of the community develop amusing reactions to the
failures of the SMoC.  A prominent Cambridge member of the
observational community searching for new satellite galaxies around
the MW at the IoA simply said, after my colloquium in June 2013, that
the VPOS (described in much detail by \cite{MKJ07, MKL08,
  Pawlowski12a, Pawlowski13}) simply does not exist.  Here it is to be
noted that the new discoveries (since 2012) of satellites beyond about
10~kpc distance achieved by non-Cambridge teams showed that each and
every newly discovered system lies {\it in the VPOS}
\cite{Pawlowski14b, KJ14}.  Until now there is no evidence for any
object beyond about 10~kpc from the centre of the MW which does not
seem to be linked to the VPOS. Despite the research papers by Metz et
al. and Pawlowski et al. being the foundational research papers on the
properties of the VPOS and the amazingly organized and symmetrical
structure of the Local Group (in contrast to the chaotic but
spheroidal structure expected from the SMoC merger tree, see
Fig.~\ref{fig_pk:LG}) \cite{Pawlowski13b}, their papers are
systematically not being cited by some of the prominent teams who
explicitly work and publish on the Local Group, its structure and
properties and on the spatial distribution of the satellite galaxies
of the MW and of Andromeda (e.g. \cite{HWP14, BEB13, BEB14,
  Diaz14})\footnote{The titles of the research papers in question are
  \cite{HWP14}: ``Too Many, Too Few, or Just Right?  The Predicted
  Number and Distribution of Milky Way Dwarf Galaxies''; \cite{BEB13}:
  ``Triaxial cosmological haloes and the disc of satellites'';
  \cite{BEB14}: ``On Asymmetric Distributions of Satellite Galaxies'';
  \cite{Diaz14}: ``Balancing mass and momentum in the Local Group''.
}. The point here is not that individual papers are not cited, but
that there appears to be a systematic attempt by researchers at some
prestigious institutions at possibly suppressing or ignoring some
rather important research results.\footnote{Citing from the A\&A
  author’s guide - July 2013: ``Papers published in A\&A should cite
  previously published papers that are directly relevant to the
  results being presented. Improper attribution -- i.e., the
  deliberate refusal to cite prior, corroborating, or contradicting
  results -- represents an ethical breach comparable to plagiarism.''}
Obviously this unscholarly behavior may have an effect on young
researchers which may be susceptible to non-scientific influence.

The way in which it is being argued, perhaps too often, on the
scientific platform may seem, at least to some, problematical: The
VPOS can neither have been formed from the accretion of a group of
dwarf galaxies \cite{Metz09} nor from the accretion of satellite
galaxies onto the MW, which is a 10-11~Gyr old disk galaxy, along a
cosmological filament \cite{Pawlowski12b}. There is therefore no
solution for the DoS/VPOS in terms of dark-matter satellite galaxies,
such that \cite{Libeskind11} suggest, ``While the planarity of MW
satellites is no longer deemed a threat to the standard model, its
origin has eroded a definitive understanding.'' At a topical meeting
of the American Astronomical Society in 2013 a prominent astronomer
declared that the Local Group cannot be used to test the SMoC. In the
end, perhaps the MW and even the Local Group may just be extreme
outliers, exceptions as it were.\footnote{This argument is invalid as
  long as properties are studied which are generic such as the wide
  (on scales of $>100$s of kpc) distribution of matter and its
  correlation in phase-space, which is not sensitive to the details of
  sub-grid baryonic physical processes.  Testing the SMoC against
  observational data becomes inconclusive if specific questions are
  raised such as where a particular satellite galaxy is, what the
  number of satellites is and which star formation history a
  particular galaxy may have had.}

But, meanwhile more trouble for the SMoC emerged: Andromeda also has
an extreme version of a rotational DoS! This shocking result has been
established by the seminal work of Ibata et al. \cite{Ibata13,
  Ibata14}. A brilliant idea of testing how often phase-space
correlated satellite systems occur was conceived and applied by the
same remarkable team \cite{Ibata14b}. They showed with very high
statistical significance that satellites on opposite sides of their
host galaxies and within projected distances of about 100--150~kpc,
have anti-correlated line-of-sight velocities in contrast to no
correlation expected if the satellites are dark-matter dominated PDGs,
as explicitly tested using SMoC simulations and intuitively expected
because PDGs merge with host dark mater halos more or less
isotropically and stochastically, and most importantly, independently
of each other.  This is independently supported by the known satellite
systems around a number of well observed host galaxies also having
phase-space correlated satellite systems \cite{Pawlowski14b}.
\begin{svgraybox}
  A major host galaxy thus appears to typically have its faint
  satellite galaxies arranged in disk-like rotating structures within
  distances of 100-150~kpc, at odds with the expectations from SMoC
  models.  The conclusion by \cite{Chiboucas13} ``In review, in the
  few instances around nearby major galaxies where we have
  information, in every case there is evidence that gas poor
  companions lie in flattened distributions.''  is noteworthy here.
\end{svgraybox}

Despite the observational evidence for rotational DoSs being so strong
and the robustly obtained result using various statistical tests that
such highly phase-space correlated satellite galaxy populations do not
occur in the SMoC if these are composed of dark matter dominated PDGs
(many research papers by Metz et al., Pawlowski et al., Ibata et al.),
various teams nevertheless attempt at demonstrating that this is not
so. Thus \cite{GB13} (until now not published by a journal, but also
not withdrawn) argue that such phase-space correlated structures arise
naturally in the SMoC if satellite galaxies are accreted along thin
baryonic streams. This argument has never been shown to actually work
in any simulation, and is ruled out since dark matter satellites will
not line-up like beads on a string as they fall into the major host
halo \cite{Kroupa14}. \cite{Pawlowski14} have ruled out this
beads-on-the-string scenario. In another attempt, \cite{BB14} retest
the claims by \cite{Ibata13} apparently discovering that, contrary to
the deduction by \cite{Ibata13}, SMoC models do naturally yield DoSs
as found around Andromeda. \cite{Ibata14, Pawlowski14} showed that
their conclusions to be flawed because they did not apply the
appropriate criteria (e.g., searching for disks using criteria that
also allow spheres will yield the wrong conclusion on the frequency of
occurrence of disks).  {\it Nevertheless, various teams cite these flawed
papers (e.g. \cite{BEB14}) without citing the rebuttals nor the
phase-space correlated DoSs.}

The most recent and amusing development in this arena is the rebuttal
by Cautun et al.  \cite{Cautun14} of the Ibata et al. \cite{Ibata14b}
result concerning the overabundance of anti-correlated satellite pairs
when the real galaxies are compared to SMoC galaxies. \cite{Cautun14}
neither show that \cite{Ibata14b} have made an error nor do they
demonstrate that the analysis of \cite{Ibata14b} is flawed, they
merely argue that the significance of the result obtained by
\cite{Ibata14b} weakens significantly when the selection criteria are
changed, e.g. by using a larger search radius around the host
galaxies. Ibata et al. \cite{Ibata14c} have already countered
this. Essentially, phase-space correlated satellite galaxy systems,
such as the DoSs of the MW and of Andromeda, do have physically
relevant scales and these are similar for major host galaxies
(100--200~kpc).

Thus, none of the attempts (\cite{GB13, BB14, Cautun14}) to argue away
that the DoSs are in highly significant disagreement with the
expectations form the SMoC models, have added much to progress, but
these attempts are cite-able research papers which make-believe that
the original, rigorous discovery papers may not be as solid. The
contents of the introductions to, e.g.  \cite{BEB14, Cautun14}, are
cases in point. Speaking to individual researchers, the sense seems to
have arisen that {\it opinion} is what counts, rather than hard
evidence: ``Yes, but they are voicing a different opinion, so the
situation is not clear.''  Perhaps this is why such weak counter
arguments, which are even partially flawed, get published at all; the
authors are allowed to voice their opinion more than actually showing
that another analysis is faulty. Thus, \cite{Cautun14} might have
rather written that their analysis agrees and supports the original
paper by \cite{Ibata14b} on the anti-correlated satellites, which it
does. They may then have shown with their analysis that the signal
disappears beyond a certain radius scale, which would have been an
interesting new result.  Therewith they could have constructively
pointed out the new valuable scientific result that
phase-space-correlated satellite systems occur only within a limited
radial scale about their host galaxies. Instead, the stance is taken
to try to weaken the seminal results by \cite{Ibata14b}. Finally I
would like to point out why I have the impression that the scientific
method may indeed be already compromised: in~2009 Libeskind et
al. \cite{Libeskind09} and in~2011 Deason et al. \cite{Deason11} make
very specific, explicit and valuable statements on how likely DoSs are
around MW like galaxies if the SMoC is assumed to be valid, given that
$N$ satellites have measured proper motions that make them
co-orbit. From Pawlowski \& Kroupa (2013, \cite{Pawlowski13} and as
re-explained in the Introduction of \cite{Pawlowski14}, the current data
robustly rule out the SMoC models, given the results of the SMoC teams
\cite{Libeskind09, Deason11}:
\begin{svgraybox}
  At least 6 (and up to 8 within current uncertainties) out of the
  brightest 11 of the real satellites have orbital angular momenta
  aligned within 22 degrees of their average angular momentum
  direction which has a probability of 0~per cent according to fig.9
  of \cite{Libeskind09}.  With this test alone the SMoC is ruled out,
  given the data.  {\it The perplexing issue however is that the
    community opinion appears to be that this is not the case},
  without having ever shown that the work of \cite{Libeskind09,
    Deason11, Pawlowski14} may be wrong.
\end{svgraybox}
The community appears to have developed an unhealthy sense of simply
ignoring or burying previously obtained results if these are highly
inconsistent with the SMoC. The above failure is terminal, that is, no
baryonic physics or other effects can change the results. Herewith the
SMoC is ruled out as a viable theory of the Universe.  \footnote{ A
  particularly troubling case in point is the manuscript published on
  the archive by \cite{Sawala14} in which they claim that by adding
  baryons to the structure formation simulations the ``missing
  satellite'' problem, the ``too big to fail'' problem and the
  ``planes of satellites'' problem are solved. I have been receiving
  e-mails from members of the community pointing this out. A detailed
  analysis of this manuscript reveals that their fig.~1 is made with a
  radiative transfer code which gets processed significantly before
  being made into the figure such that the faintest features in the
  figure are unknown. The manuscript thus contains material which
  cannot be quantified. Furthermore, the authors did not test the
  planarity of the 3D distribution, but use only two dimensions. They
  do not test for the clustering of orbital angular momentum poles of
  their model satellites and their data are in fact perfectly
  consistent with a purely random spheroidal dark-matter-only
  simulation. There are other flaws, e.g. concerning their test of the
  Andromeda system, and generally it can be stated that they re-phrase
  the known problems in a way such that they can pretend to have
  solved them using not very robust tests or no real tests.}

In addition to a certain degree of perhaps dis-honestness of parts of
the cosmological community alluded to above, the unhealthiness of the
cosmology-relevant research environment is suggested with the
following few instances: A close collaborator of mine, with whom I
began the project which ultimately resulted in the publication
\cite{Kroupa10}, dropped authorship of this paper because of the fear
that having his or her name on this paper would compromise his or her
chances of obtaining a job in astrophysics. Two other very close
collaborators of mine refuse to cite \cite{Kroupa12a} and
\cite{Kroupa14}, despite being relevant for the arguments in their own
papers, again, because citations to these two publications of mine may
mire their own papers in the eyes of the community. I know of some
authors being told by referees that citations to these two papers of
mine should be dropped.  I have spoken to fairly senior scientists who
refuse to openly criticize the SMoC for fear of having their chances
reduced for obtaining research grants. Especially in the USA this is
problematical because the summer salary depends on obtaining grants in
a highly competitive environment. A brave PhD student, on giving a
presentation on MOND in Cambridge, was asked publicly by a highly
senior astronomer how many angels are on a pin head. A PhD student of
mine was told, again by someone from Cambridge, that he or she ought
to stop following me, and I have also been told by a very senior
astronomer that I should stop arguing that there can be no dark
matter, and a German die-hard SMoC director walked out of one of my
colloquia at his institution in northern Germany. Instances are
documented where attempts are made in removing the research budget and
secretarial support from ``dissidents'', whereby other reasons than
scientific ones need to be applied.  Clearly, if legal action were to
be taken on such instances, a scandal may ensue.  Apart from hindering
an advancement in understanding of data (such as the above mentioned
notion that the Local Group data cannot be used to test the SMoC),
such incidences contribute to a general atmosphere of fear of
criticizing the SMoC.
\begin{svgraybox}
\label{pk:misconduct}
  The above examples are uncomfortable but reveal that the research
  environment concerning cosmology has become unhealthy. Too many
  researchers are willing to write flawed manuscripts but are not
  willing to make significant conclusions when their results
  contradict or exclude the SMoC. This is fully consistent with the
  adverse behavior noted in other scientific disciplines
  \cite{Corredoira09,Fanelli10,TheEconomist13a,TheEconomist13b}
  and may be due to the ``perish or publish'' doctrine married to a
  strong bias towards seeing the SMoC as the only valid model which
  was established by the now highly influential senior scientists.
\end{svgraybox}

\subsubsection{Implications and a comment on history}
\label{sec_pk:impl}

If dark matter particles do not exist then there are a number of
implications. Some are listed here, whereas the more fundamental one
is discussed in Sec.~\ref{sec_pk:conseq}.

\begin{itemize}

\item The direct implication for the galaxy population is that {\it
    mergers between galaxies are very rare in reality}, because the
  dynamical friction on the expansive and massive dark matter halos
  does not occur, as already realized by Toomre in~1977
  \cite{Toomre77}.  This is further evident in the ratio of the number
  of star-forming galaxies over elliptical galaxies ($\approx\,$a~few
  percent) not changing within the uncertainties since redshift
  $z\approx6$ (fig.~7 in \cite{Conselice12}).

\item The large observed uniformity of the galaxy
  population with more than 90~per cent of all galaxies with
  luminosity $L>1.5\times 10^{10}\,L_\odot$ \cite{Delgado10} being
  disk galaxies which are on the main sequence \cite{Speagle14}
  then becomes readily understandable. This is because the haphazard,
  stochastic and different merger histories which would lead to a
  large variation of galaxy properties \cite{Disney08}, does not
  occur. Instead, galaxies are self-regulated \cite{Koeppen95} slowly
  growing, largely isolated structures which sometimes interact, such
  as the MW and Andromeda \cite{Zhao13}. During encounters dwarf
  galaxies are formed as TDGs in DoSs.

\item Understanding galaxies purely as baryonic, self-gravitating
  systems becomes simple and predictive with remarkable success: That
  one disk galaxy resembles another follows from self-regulation, as
  just noted above. Taking a dark-matter-free TDG, placing it on an
  orbit within the potential of a host galaxy, leads to surprising
  agreement with the observed satellite galaxies of the MW and of
  Andromeda. Before delving into this, I'd like to explain how I
  managed to stumble onto the result that the SMoC is not valid
  (Eq.~\ref{eq_pk:fals}):

  {\small I have had no reason to be against dark matter, and indeed
    in the history of physics there were incidences when hitherto
    unknown particles or bodies were postulated to be present which
    were also found. Famous examples are the neutrino and the planet
    Neptune. The brilliant work of Einstein and the discovery of
    redshifted galaxies were beautifully consistent, as was the
    discovery of the cosmic microwave background radiation flux with
    its remarkable properties which support the notion of a hot Big
    Bang. And so, in order to make structure formation work in an
    inflated and thus diluted early universe and to account for the
    high dynamical mass-to-light ($M/L$) ratios of some galaxies, the
    postulate that there must be cold or warm particle dark matter
    which is dynamically and gravitationally dominant on the scales of
    galaxies was to be taken very seriously indeed
    (Eq.~\ref{eq_pk:theSMoC}). Because the equations of motion in the
    appropriate limits remained linear and Newtonian, it was easy to
    develop computer programs to calculate the emergence and the
    formation of structure in the SMoC universe. With increasing
    computer power the calculations became relevant for studying the
    properties of the computer galaxies, and I remember how convincing
    the movies of disk galaxies with their satellite galaxies looked
    like in the mid-1990's, as shown for example in a brilliant
    colloquium held in Heidelberg by Simon White. I remember this very
    vividly indeed. A very strong, extremely well funded international
    community developed in many countries performing such
    computations. With the realisation that the computations did not
    yield the observed galaxies, the many world-wide active groups
    became entrenched in ever more conflicts concerning fine tuned and
    detailed considerations of sub-grid baryonic physics processes.}

  {\small During the early 1990's I have been witnessing these
    developments as a spectator, as at that time I was much more
    concerned in understanding how populations of binary stars in star
    forming regions, which had been observed for the first time with
    the new infrared observing platforms, could be made consistent
    with the significantly lower binary fraction in the Milky Way
    field population. This is where I learned stellar dynamics,
    essentially by reading and from Sverre Aarseth in
    Cambridge. Sverre is perhaps the most remarkable and stimulating
    scientist I have come to know, and he influenced me tremendously,
    probably more than any other person. Only decades later in~2011
    did I learn that Sverre had actually done, in~1979, the very first
    and pioneering cosmological structure formation simulation
    \cite{ATG79}, but he had left this too-speculative field
    \cite{Kroupa12a}. In the early 1990s I lived in Heidelberg but,
    still being a Rouse Ball Scholar, visited him often, bringing with
    me a good supply of German white wine, in expectation of receiving
    driving lessons with his Nbody codes. At that time he refused to
    do sacrilegious experiments with young star clusters containing
    100~\% binaries, since at that time it was thought that only a few
    percent of binaries are in clusters. I had to do these
    computations myself, and for that the array structures in his
    codes had to be updated. In any case, then the binary-star problem
    was a hobby\label{p_pk:hobby}, as at that time I was employed in
    Heidelberg by Roland Wielen to work on satellite galaxies and
    their dynamical and possibly disruptive influence on the thin
    disks of major spiral galaxies. I had no reason to consider that
    dark matter did not exist, and I began this work by first learning
    to set up spherical satellite galaxies and computing their
    evolution in the dark matter halo of the Milky Way. I used the
    programme {\sc Superbox} developed by Reinhold Bien \cite{Bien91,
      MB93, Fellhauer00} in Heidelberg, a particle-mesh code with
    nested sub grids. After some adjustments and modernisations of
    this code I obtained my first solutions which had me raise my
    eyebrows.}

  What I found did not fit into the general picture as portrayed by
  the community that the properties of the satellite galaxies of the
  Milky Way can only be understood if they had their own substantial
  dark matter halos. They were observed to have very large dynamical
  mass to light ratios, larger than ten, and in some cases a few
  hundred. My own computations yielded hitherto unknown solutions to
  this finding \cite{Kroupa97}.

  My satellite models assumed that mass-follows-light, they were
  set-up as TDGs, and I calculated their evolution in very high
  resolution for that time for a Hubble time and on eccentric orbits
  within the massive and extended dark matter halo of the Milky Way. I
  found the satellites to loose particles at every periastron, until,
  after only about 10--1~per cent particles remained in them, a sudden
  change in their properties occurred. They catastrophically lost even
  more particles, but a remnant always remained for many more
  orbits. These remnants, containing about 1~per cent of their
  original particles, were quasi stable and clearly identifiable
  condensations of particles in phase space and in orbit about the
  Milky Way model. Observing them in the computer as a real observer
  does real satellites, by removing velocity outliers, I found these
  remnants to have tremendously high dynamical (apparent, i.e. not
  true) $M/L$ ratios and velocity dispersion and spatial extensions
  comparable to the real satellite galaxies. I published these results
  in a paper with the title ``{\it Dwarf spheroidal satellite galaxies
    without dark matter}'' \cite{Kroupa97}.\footnote{This paper, by
    the way, drew interesting comments by a few well-established
    senior scientists. Once, in a train going back from Switzerland
    from a conference, I was told that writing a paper with such a
    title is suicidal. At a dinner in 2003 a very senior professor
    told me in Bonn that I am un-hirable by having written such a
    paper.} This work demonstrated to me that what we are observing as
  real satellite galaxies may not be dark-matter dominated objects
  after all. Further work confirmed these results using an entirely
  different computer code \cite{KK98} (see also \cite{Casas12} for a
  parameter survey). With PhD student Manuel Metz we reconsidered the
  observational properties of these systems, finding them to be
  surprisingly similar to the real satellite galaxies
  (Fig.~\ref{fig_pk:hlr}).
\begin{figure*}
\centerline{
  \resizebox{18cm}{!}{
   \includegraphics{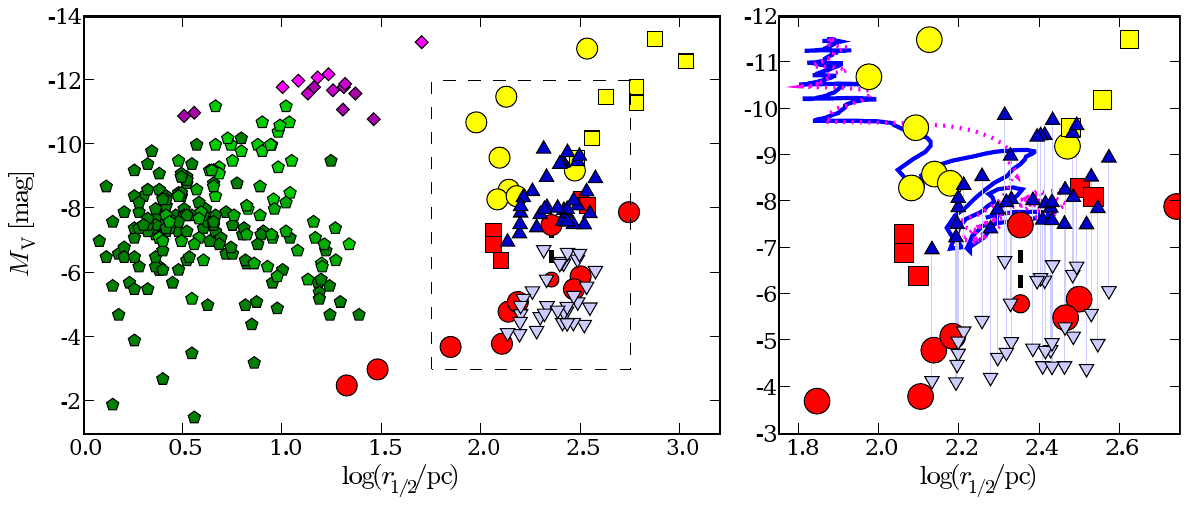}
 }}
\vspace{-36mm}
  \caption{ Absolute magnitude versus half-light radius for Globular
    Clusters (pentagrams), ultra compact dwarf galaxies (diamonds),
    Milky Way (circles) and Andromeda (squares) dSph satellite
    galaxies, and simulated ancient satellite TDGs (triangles). For
    the observational data of the MW and Andromeda, the dark green
    symbols mark the ultra-faint dwarf (UFD) additions until 2007 to
    the known list of companions from the SDSS, while light yellow
    symbols mark the longer-known (``classical'') dSph's. For the MW
    dSph galaxy in Bo\"otes two values for the absolute magnitude are
    given (see \cite{MK07}), the values are connected by a thick,
    dashed line. For the simulated data by \cite{MK07} absolute
    magnitudes derived within a fixed projected distance $r_{\rm
      bin}=1.5\,$kpc (dark triangles) and within the variable
    half-mass radius $r_{1/2}$ (light triangles) are shown.  No
    parameter adjustments are applied; mass-follows light and the
    adopted V-band mass-to-light ratio of each particle is
    $\left(M/L\right)_{\rm true}=3$.  In the right panel, the region
    in the left panel marked by the dashed rectangle is
    enlarged. Corresponding absolute magnitude values for the
    simulated satellites are connected by light solid lines. The
    evolutionary tracks of the models RS1-5 (solid curve) and RS1-113
    (dotted curve) are shown. Adapted from \cite{MK07}, their fig.~2.
    The models have observed, dynamical $M/L > 100$ for most cases
    (fig.~3 in \cite{MK07}). }
  \label{fig_pk:hlr}
\end{figure*}

Particularly stunning, in retrospective, was the accidental prediction of a
Hercules-type satellite galaxy I made in the 1997 paper: 
model RS1-5 of \cite{Kroupa97}, shown in that paper 
as a snapshot (and here in Fig.~\ref{fig_pk:hercules}), is an essentially perfect
match to the dSph satellite Hercules (see fig.~2 in \cite{Coleman07})
discovered~10 years later by \cite{Belokurov07}.  The half-light
radius is 180~pc in the model and 168~pc for Hercules, RS1-5 has a
velocity dispersion of about 2.8~km\,s$^{-1}$ (table~2 in
\cite{Kroupa97}), while Hercules has a measured velocity dispersion
of $3.72\pm0.91$~km\,s$^{-1}$ \cite{Aden09}, and the inferred
mass-to-light ratio that one would deduce from velocity dispersion
measurements based on the assumption of equilibrium is about $200$ in
both cases.  Both RS1-5 and Hercules have luminosities agreeing within one
order of magnitude (the model being the brighter one), yet RS1-5 has
no DM. 
\begin{figure*}
\centerline{
  \resizebox{18cm}{!}{
   \includegraphics{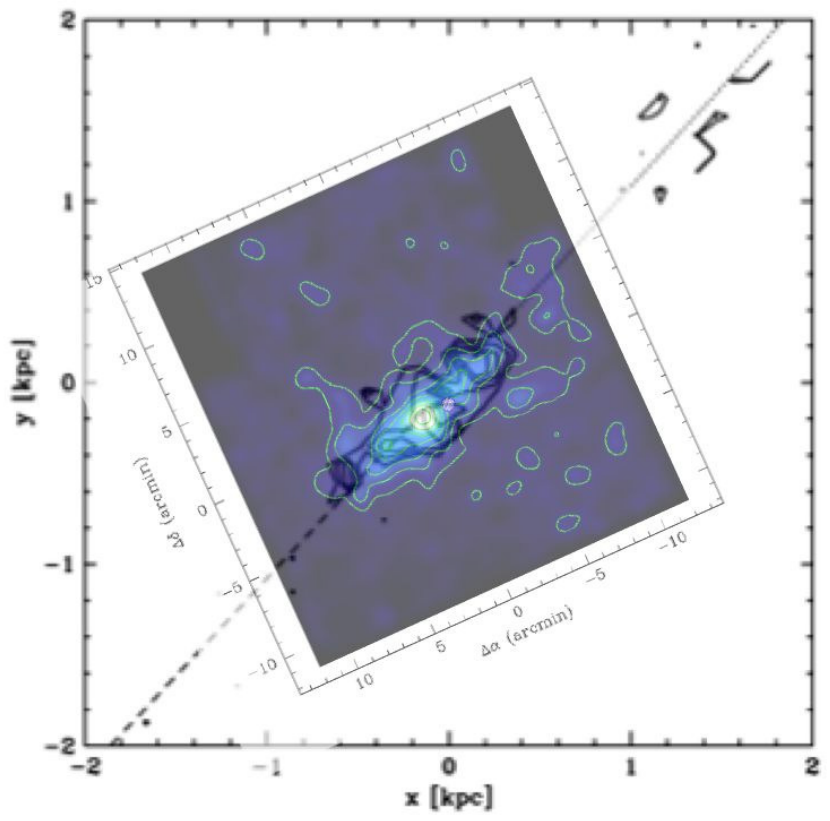}
 }}
\vspace{-18mm}
\caption{Model RS1-5 from \cite{Kroupa97} (on the kpc grid) is plotted
  over the surface brightness contours of Hercules by \cite{Coleman07}
  (celestial coordinate grid). The dashed and dotted curve are,
  respectively, the past and future orbit of RS1-5.
\label{fig_pk:hercules}}
\end{figure*}

I am not aware of any other (dark matter based) models having been
able to {\it predict} with such accuracy a really found satellite
galaxy (Fig.~\ref{fig_pk:hercules}). I am also not aware of any other
models which, without fine-tuning of parameters, so naturally account
for the observed properties of the MW and Andromeda satellites
(Fig.~\ref{fig_pk:hlr}).\footnote{I should emphasize that today I do
  not consider these solutions to be valid in their entirety:
  If there is no dark matter in galaxies, then the flat rotation
  curves of disk galaxies force us to use effective scale-invariant
  dynamics (SID, or Milgromian dynamics). The computations I published
  in 1997 \cite{Kroupa97} thus need to be redone in this gravitational
  framework. More on this in Sec.~\ref{sec_pk:conseq}.  }

{\small At this time (late 90s) I could have left this field, but I
  sensed that this issue of the satellites may well be central to
  uncovering a fundamental property of the Milky Way, and more
  importantly, may impinge on the issue of the existence of dark
  matter. Thus I began to seek other clues, and the spatial
  distribution of satellite galaxies around the Milky Way was such a
  clue. It was well known from the much earlier work in the 1970s by
  Donald Lynden-Bell \cite{LyndenBell76, LyndenBell82} and Kunkel \&
  Demers \cite{Kunkel76} that the satellite galaxies and the distant
  globular clusters were in streams. But their work was either ignored
  or forgotten by the cosmological community which needed the
  satellite galaxies to be part of the dark-matter dominated sub-halo
  population. Indeed, the satellite galaxies do have very large
  dynamical $M/L$ ratios (if the analysis is done assuming Newtonian
  dynamics), so this was a logically correct association. However, my
  solutions published in 1997 \cite{Kroupa97} suggested that this
  interpretation with dark matter may not be needed. So the highly
  anisotropic distribution of the then known satellite galaxies became
  interesting and telling. My own tests in~2005 with Christian Boily
  and Christian Theis of the real distribution against that expected
  from the haphazard merging of sub-halos due to the cosmological
  merger tree showed that the latter was inconsistent with the real
  spatial distribution with a high significance \cite{Kroupa05}.  With
  my PhD students Manuel Metz and Marcel Pawlowski we continued this
  work using actual SMoC simulation data as the null hypothesis and
  more sophisticated statistical tests, some of which were ported from
  the mathematical literature, and we followed the new discoveries:
  initially we were excited with each new satellite galaxy, and even
  more thrilled when it turned out that every new discovery up until
  now strengthened the DoS (more generally the vast polar structure,
  VPOS) \cite{MKJ07,Pawlowski12a, Pawlowski14b, KJ14}, that it is
  rotational \cite{MKL08, Pawlowski13} and that it is an extremely
  robust structure defying an explanation within the SMoC
  \cite{Metz09,Pawlowski12b,Pawlowski14, Ibata14, Ibata14c}, that its
  explanation as mapping out fossil tidal arms is consistent with them
  being TDGs \cite{MK07, Pawlowski11}, and that the entire Local Group
  appears to be highly structured and symmetric
  \cite{Pawlowski13b}. Other such systems have begun appearing too
  \cite{Chiboucas13,Pawlowski14b, Ibata14b}.}

In this context, since 2007 the work of Gerhard Hensler and his group
in Vienna on theoretically understanding the formation and evolution
of dark-matter free TDGs is impressive and ground-breaking: Recchi et
al. in 2007 \cite{Recchi07} and Pl\"ockinger et al. in 2014 \& 2015
\cite{Ploeckinger14,Ploeckinger15} are significantly contributing to
our understanding of self-regulated star-forming TDGs showing them to
follow a mass--metallicity relation and to survive for many Gyr
despite star-formation and baryonic feedback processes. That TDGs
emerge very naturally in gas-rich galaxy--galaxy encounters has
already been demonstrated by the pioneering work \cite{Wetzstein07,
  Hammer10, Fouquet12, Yang14}. This enhances the missing satellites problem of the
SMoC because TDGs without dark matter ought to be abundant according
to these simulations, all of which were made within the SMoC
framework. That the number od TDGs produced within the SMoC is as
large as the observed number of dE and other satellite galaxies has
been pointed out by Okazaki \& Taniguchi \cite{OT00}. 

\item The existence of the main sequence of galaxies at various
  redshifts \cite{Speagle14, Tasca15} is remarkable and has not been
  predicted within the SMoC. Indeed, the apparently highly
  self-similar behavior of galaxies of different masses in terms of
  their star formation rates being simple functions of their stellar
  masses remains unexplained, and intuitively unexplainable by any
  stochastic haphazard merger-driven galaxy evolution theory
  \cite{Disney08}. The notion that the main sequence is a consequence
  of smooth accretion onto the galaxies may be correct, but how is it
  to be arranged given the cosmological merger tree?  Instead, the
  main sequence of galaxies appears to be more consistent with
  galaxies which are largely isolated and self-regulated objects such
  that the mergers driven by dynamical friction on the expansive and
  massive dark matter halos do not play a role.

\item Another, more fundamentally important implication is that the
  standard model of particle physics (SMoPP) therewith remains valid
  for all matter in the universe; an extension of it to accommodate
  additional (exotic) dark matter particles is unnecessary and in fact
  there is no room for such particles given the astronomical evidence,
  subject to the caveat in Sec.~\ref{sec_pk:conseq}.

\end{itemize}

\begin{svgraybox}
  In summary: Without particle dark matter theoretical galactic
  astrophysics becomes simple and predictive with remarkable success.
  Galaxies merge rarely and typically evolve as self-regulated systems
  and most satellite galaxies are TDGs.
\end{svgraybox}

\subsection{DM Question 2}
\label{sec_pk:conseq}

{\it If, as you say, the DM paradigm is wrong, what changes are necessary
to reconcile observations with theoretical models}?

\subsubsection{Answer:}
\label{sec_pk:Milgrom}

The implications of Eq.~\ref{eq_pk:fals} are manyfold, and some of
the more obvious astrophysical ones are discussed in
Sec.~\ref{sec_pk:impl}. 
\begin{svgraybox}
  If there is no cold or warm dark matter particle, then particle
  physics is in a very good shape. But then, effective gravitation
  needs to be non-Newtonian.
\end{svgraybox}
``Effective'' gravity means that either Einstein's field equation
needs to be extended (e.g. \cite{Bekenstein04}, see
also\cite{Moffat06}), or it remains valid but processes in the vacuum
may change the equations of motion \cite{Milgrom99}.

A promising avenue for further research is, to my mind, the
implications of the following scaling of Minkowski space-time:
$(\vec{r}, t) \longrightarrow \lambda(\vec{r}, t)$, where $\vec{r}$ is
a Cartesian vector and $t$ the time and $\lambda$ a number.  Demanding
that the gravitational acceleration scales as $a_g \longrightarrow
\lambda^{-1} a_g$, just as the kinematical acceleration does,
immediately leads to basically all observed laws of galactic dynamics
in the weak gravitational regime. The essence here is that this last
requirement yields (for spherical systems) $a_g =
\left(a_0\,a_N\right)^{1/2}$, where $a_N$ is the Newtonian
gravitational acceleration and $a_0$ is Milgrom's constant which needs
to be introduced simply on dimensionality grounds. A single galaxy
rotation curve yields $a_0\approx 3.8\,$pc/Myr$^2$, and realms within
which the acceleration $a<a_0$ are the weak-field regime.  This was
pointed out by Milgrom in~2009 \cite{Milgrom09} and discussed
didactically by \cite{Kroupa14} and \cite{WK15}.

These latter discussions clarify how a Newtonian observer will
misinterpret the motions of satellite galaxies or of matter in disk
galaxies as being due to an isothermal dark matter halo. This halo is
not real, it is merely a mathematical formulation when the gravitation
being used is Newtonian. In Milgromian, or scale-invariant dynamics
(SID), there is no particle dark matter halo. Galaxies effectively
exert a stronger gravitational force on their neighbours, because
$\left(a_0\,a_N\right)^{1/2} > a_N$ for $a_N < a_0$, such that the
equivalence principle is broken. 
\begin{svgraybox}
  That is, a galaxy subject to an external force will accelerate
  according to its baryonic mass only, but it will pull at other
  galaxies with an effective larger force, as if it were heavier
  through a phantom dark matter halo.
\end{svgraybox}
The phantom halos are effectively truncated through external fields,
and this leads to a new gravitational phenomenon unknown to Newtonian
dynamics. The {\it external field effect} (EFE) makes Milgromian
gravitation richer and may well explain a few hitherto unaccounted for
observations \cite{Kroupa14, WK15}.  A transition into the Newtonian
regime when the acceleration is larger than Milgrom's constant,
$a_N>a_0$, can be deduced from processes in the vacuum, as suggested
by Milgrom \cite{Milgrom99}, or simply by designing a transition
function $\mu$. This is an equivalent process to how Max Planck
originally introduce Planck's constant prior to understanding its role
in quantum mechanics.  The whole mathematical description
incorporating the Newtonian regime and the SID regime, is known as
Modified Newtonian Dynamics, or MOND, and has been extensively
discussed and originally introduced by Milgrom in~1983
\cite{Milgrom83}.

MOND is a classical effective but non-linear theory of gravitation
with a generalized Poisson equation (eq.~3 in \cite{BM84}).  Attempts
have been made in developing general relativistic formulations (for
the break-through paper see Bekenstein's 2004 \cite{Bekenstein04}, see
also \cite{Bekenstein11}). Reviews of this entire subject are
available in Famaey \& McGaugh (2012, \cite{FM12}) and Trippe (2014,
\cite{Trippe14}), see also the references provided in \cite{Kroupa14}.

Without dark matter the SMoC is ruled out as a relevant model for the
Universe. 
\begin{svgraybox}
  This means that the redshift-distance, redshift-age relations are
  likely to be different, thus much of what we know about the Universe
  from observations may need perhaps major revision.
\end{svgraybox}
First steps towards developing cosmological models which are
consistent with SID have been undertaken by the seminal work of
Claudio Llinares \cite{Llinares08} and Garry Angus
(e.g. \cite{Angus13}). But such simulations still need to rely on an
assumed Einsteinian expansion history of the Universe which is the
same as that of the SMoC, the simulations treat the baryons as
non-dissipative non-star-forming particles which thus essentially
behave just like the dark matter particles of the SMoC, and a hot dark
matter component in the form of, for example, 11~eV sterile neutrinos,
needs to be postulated (e.g. \cite{Angus09}). This hot dark matter
component may be less of a problem than the exotic cold or warm dark
matter particles required in the SMoC, because it may be added to the
SMoPP naturally in accounting for the masses of the active
neutrinos.\footnote{Note that the often invoked limits on neutrino
  masses using standard-cosmological arguments become invalid by
  discarding the SMoC.}

Thus, a very conservative cosmological model appears to be emerging
\cite{Kroupa14}: it is based on GR, has inflation and dark energy, and
the exotic cold or warm dark matter particles of the SMoC are
naturally avoided by vacuum effects yielding SID with sterile
neutrinos entering as a hot dark matter component. This model has been
introduced by Garry Angus, who constraints the mass of the sterile
neutrino to be about 11~eV \cite{Angus09, Angus10, AD11, Angus13}.
As of a few months ago full-scale purely hydrodynamic simulations of
structure formation in MOND have become possible with the code {\it
  Phantom of Ramses}  (PoR) developed by Fabian L\"ughausen in
collaboration with Benoit Famaey and myself as a patch to 
Romain Teyssier's {\sc RAMSES} \cite{Lueghausen14b}. 

{\it A few final comments}: Given that the SMoC is ruled out, new
models are being studied which naturally account for the observed
gravitational and dynamical properties of galaxies without dark
matter. However, given the pressure in the community against non-SMoC
research, progress is very slow and even haltering. The widely-held
interpretation that the population of galaxies we observed today is a
result of mergers is certainly wrong; mergers play a minor role only
(Sec.~\ref{sec_pk:impl}).  {\it Galaxies, as they are observed, cannot
  be reproduced in the SMoC.}  Observational cosmology is providing
important new constraints, but applications of the redshift--distance
and redshift--age relations from the SMoC to interpret these data is
almost certainly wrong in that {\it the true nature of the physical
  systems at high redshift is likely to be distorted when interpreted
  within the framework of the SMoC.}

\section{The stellar initial mass function (IMF)}

The Initial Mass Function (IMF, $\xi(m)$) is one of the most important
theoretical ingredients of any theory of galaxy formation and
evolution.  The concept of the IMF was first introduced by
\cite{Salpeter55}.  The IMF is defined to be the differential number
of stars, $dN$, in the stellar mass interval $m$ to $m+dm$,
$dN=\xi(m)\,dm$. It is the distribution function of all stars formed
together in one ``event'', and the Salpeter IMF \cite{Salpeter55} is
$\xi(m)\propto m^{-\alpha}, \alpha\approx 2.3, 0.4<m/M_\odot<10$.  It
provides a convenient way of parametrization of the relative number of
stars as a function of their mass. The IMF has been one of the most
debated issues in galaxy studies.  
\begin{svgraybox}
\label{p_pk:IMF}
Modern measurements obtained from young clusters and associations, and
old globular clusters suggested that the vast majority of their stars
were drawn from a universal or canonical IMF: a power law of Salpeter
index ($\Gamma_2 = 1.3, \alpha_2=2.3$) above halve a solar mass, and a
log normal or a shallower power law ($\Gamma_1 \approx 0.3, \alpha_1
\approx 1.3$) for lower mass stars (fig.~4-24 in the recent review
\cite{Kroupa13} which covers much of this material).
\end{svgraybox}

\subsection{IMF Question 1}
{\it The shape and the universality of the IMF is still under investigation. Could you explain why}?

\subsubsection{Answer:}

Before continuing one needs to establish some precise vocabulary: The
stellar IMF is the distribution of stellar masses formed together in
one star-formation event (see Sec.~\ref{sec_pk:eventIMF}). It is
constrained by star counts in a given star-formation event. Such a
population of stars is {\it simple} (one age, one metallicity).  The IMF
of a whole galaxy (see Sec.~\ref{sec_pk:galIMF}) is a different issue,
as it is deduced from the field population of stars in a galaxy, and
this field population has many different ages and metallicities, it is
{\it complex}. 
\begin{svgraybox}
  Rigorous work on the IMF needs to differentiate between the true IMF
  of a simple population and the IMF of a complex population. Are the
  two the same? 
\end{svgraybox}
The reason why the question of whether the IMF is universal or not is
still being studied and debated is that the IMF is indeed such a
fundamentally important distribution function, and because
constraining the IMF observationally is very hard indeed and mistakes
in the analysis can easily occur if the work is not highly rigorous in
every respect.  Any scientist attempting this task requires intimate
knowledge of all aspects of astrophysics, such as pre-main sequence
and post-main sequence stellar evolution, stellar birth-rate
functions, the structures in which stars typically form and their
dynamical evolution including gas expulsion processes, the properties
and evolution of binary systems and the corrections of star counts for
various biases and uncertainties. One bias, for example, often not
appreciated in dealing with Galactic-field star counts, is that by the
nature of the systematically changing mass-ratio of binary stars with
primary mass, the photometric distance estimates suffer a systematic
bias in dependence of the primary star mass \cite{Kroupa93}.  It is
comparatively easy to make a survey, count the ``stars'', and to
construct a ``mass'' distribution. Following such a straight forward
procedure, typically one obtains different mass functions for
different populations (e.g. the Orion Nebula Cluster versus the
Taurus-Auriga populations vs the Galactic field ``IMF'').  But the
difficult and salient aspect of deriving the IMF is to correct the
star counts for all biases and extracting the physically relevant
information. And this is where some teams have progressed far, while
others have not, and therefore significant discussion continues.
\begin{svgraybox}
  Essentially, the problem is so hard, but appears so easy, that
  mistakes are made readily leading to debates and argumentations
  which might not be necessary.
\end{svgraybox}
Any young researcher without very detailed knowledge of all the
previous results and analyses, is likely to do avoidable and
out-of-date errors therewith setting back progress unnecessarily.

Two cases in point exemplify this: some researchers keep insisting up
until today that the IMF obtained from the Taurus-Auriga groups of
very young stars is substantially different from the normal or
canonical IMF or the IMF constrained for the Orion Nebula Cluster.
But, taking into account the very major uncertainty in estimating
stellar masses for $<$few$\,$Myr old stars (see
p.~\pageref{page_pk:Wuchterl}) and the known fact that most stars in
Taurus are in binary systems while only about 50~per cent of systems
in the Orion Nebula Cluster are binaries, leads to the underlying
parent distribution function of individual stellar masses being
consistent with the same function within the uncertainties. This has
been shown a long time ago (see the review \cite{Kroupa13}), but, for
some unclear reason, this is being ignored by others.  Another example
is the recent claim by \cite{RM13} that brown dwarfs constitute a
continuous extension of the stellar IMF based on the recently
constrained field-star IMF by \cite{Bochanski10}. But the authors of
the field-star IMF explicitly warn, in their abstract, that the
functional form of the IMF they derive is only valid for a restricted
mass range which excludes brown dwarfs. Ignoring this and using the
functional form as a model including brown dwarfs yields wrong
results. The observationally established existence of the brown dwarf
desert according to which stars rarely have brown dwarfs as companions
\cite{Dieterich12} is a primary issue where \cite{RM13} err. All of
this discussion has been occurring in the past years, although models
addressing all of these issues carefully had been published many years
ago (see the review \cite{Kroupa13}).  It has already been shown in
2003 \cite{Kroupa03} that, treating brown dwarfs as stars in
constructing binaries, leads to far too many star--brown and far
too few star--star binaries, in comparison to all known
populations. Brown dwarfs therefore absolutely must be treated with
their own, separate, IMF, as also planets have their own IMF which is
not a continuous extension of the stellar IMF.
\begin{svgraybox}
  Thus, a discussion is kept going which may not be entirely useful,
  rather than building upon the robust observational findings,
\end{svgraybox}
\noindent such as the verification of the brown dwarf desert by the excellent
work of Dieterich et al. \cite{Dieterich12} in combination with the
known stellar and brown dwarf binary properties.

\subsubsection{The universal but systematically varying IMF of simple populations}
\label{sec_pk:eventIMF}

The holy grail of IMF research is extracting the expected systematic
variation of the IMF with physical conditions of star formation. 

A star-formation event yields a stellar population whose mass
distribution is describable by the stellar IMF of a simple population
(see p.~\pageref{p_pk:IMF}).  Such a star-formation event occurs in a
molecular cloud core typically on a sub-pc-scale and on a Myr time-scale
and can be referred to as an embedded cluster. The stars belonging to
such an ``event'' can neither be counted accurately nor precisely, but
such a population is mono-metallic and coeval to within a few to ten
times $10^5$~yr, which is the time-scale over which an embedded
cluster forms. This time scale is typically a few to ten times longer
than the time ($\approx 10^5\,$yr) it takes for an individual star to
assemble about 90~per cent of its final mass \cite{WT03}.  This is
seen nicely even in supposedly ``distributed'' or ``isolated'' star
formation in the Taurus-Auriga clouds \cite{G93,B98, KB03} and in the
southern part of the L1641 Orion cloud \cite{Megeath12, Hsu12,
  Hsu13}. In these clouds the stars and proto-stars with ages younger
than about 1~Myr are distributed non-uniformly in many groups of stars
clustered on $\simless 1$~pc scales.
\begin{svgraybox}
  Thus, the direct imaging of all very young stellar and sub-stellar
  objects disprove the concept that there is a distributed mode of
  star formation below some threshold. Star formation is organized
  into sub-pc-scale events, which for all practical purposes can be
  described as embedded clusters.  Direct observations suggest that
  the least massive embedded cluster consists of about a dozen
  binaries \cite{KB03}.
\end{svgraybox}
\noindent Denser, richer embedded clusters are dynamically active and
expel stars from their cores as soon as these form (Oh, Kroupa \&
Pflamm-Altenburg 2015: \cite{OKP15}). Extremely massive star clusters
with stellar masses $\simgreat 10^6\,M_\odot$ may retain gas for long
such that their stellar populations may be complex
\cite{Wuensch11}. Even modest clusters may re-accrete gas well after
their formation \cite{PAK09} also leading to non-simple population
mixtures.Observations of the gas content of young and intermediate-age
clusters constrain such theories \cite{BastianStrader14}. 
\begin{svgraybox}
  According to the IMF un-measurability theorem \cite{Kroupa13} {\it
    the IMF can never be measured}. It can be stated that the IMF does
  not have physical reality: there is never any instant in time where
  $\xi(m)$ is fully assembled. $\xi(m)$ is therefore a theoretical and
  mathematical concept or entity.
\end{svgraybox}
\noindent As new binary stars form, others are ejected or broken up
into their binary companions, at any instant low-mass stars have not
yet reached the main sequence while massive ones have already left it
and/or have been ejected from their rich embedded clusters.  Thus,
what an observer deduces, given an available particular survey data
set, is merely a part of $\xi(m)$.

The art in the game of deducing a complete mathematical form of
$\xi(m)$ and the mass range over which it is valid (assuming such a
form exists as a theoretical construction) is putting together the
observational clues and pieces to one functional form which can be
used in theoretical work on stellar populations.  Indeed, a particular
stellar population constitutes merely a snapshot which is but
fleeting, and the same population may appear to be described by a
different mass function (MF) when viewed with a different survey at a
different (astrophysical) time. Apart from the highly significant
uncertainties (factors of two) in mass and age determinations of
individual rapidly evolving very young stars given their photometric
properties when they are younger than a few Myr, as demonstrated by
the seminal work \label{page_pk:Wuchterl} of G\"unther Wuchterl \&
Werner Tscharnuter in 2003 \cite{WT03}, there is patchy obscuration by
dust and, very fundamentally, a time-varying population of unresolved
binary stars.
\begin{svgraybox}
  Star formation typically yields binary stars, because the
  contracting pre-stellar molecular cloud core needs to shed and
  deposit its angular momentum while the formation of three- or
  higher-order multiple systems can only be a rare outcome
\cite{GK05}.
\end{svgraybox}
\noindent Rich clusters, which partially survive the violent birth
involving expulsion of their residual gas with the associated violent
revirialisation (e.g. \cite{BK13, BK14}), will, after these events,
contain a stellar mass function which has been damaged by loss of
stars and this may be stellar-mass dependent if the clusters were mass
segregated \cite{Marks08}. Star clusters evolve by evaporation
preferentially of their least-massive stars and dissolve in about
20~present-day two-body relaxation times. While the initial or
primordial binary population is broken up early \cite{MKO11}, the
binary fraction may increase with time as hard binaries\footnote{Hard
  binaries have an absolute binding energy, $E_{\rm bin}>0$, wich is
  significantly larger than the mean kinetic energy of the cluster
  stars, $E_{\rm kin}$. Soft binaries have $E_{\rm bin} \ll E_{\rm
    kin}$.} remain preferentially in a cluster because they have, on
average, higher system masses than single stars.\footnote{The issue of
  IMF invariance is related to the important issue of whether the
  initial binary-star distribution functions are invariant as
  well. Observational evidence, analysed carefully and taking into
  account the dynamical evolution properly, suggests this to be the
  case in present-day star-forming regions \cite{MKO11, MK12} and in
  major star burst clusters a Hubble time ago \cite{Leigh15}.}  Binary
systems are typically unresolvable with observations.  {\it At any
  time, a cluster thus has an observable stellar (system) mass
  function which deviates substantially from the original IMF of all
  its stars it was born with.} Direct star-formation simulations which
are already approaching sufficient realism to reflect the real
population can be used to study the time-variation of the observable
MF of stars and binary systems such as demonstrated by the seminal
work of Matthew Bate \cite{Bate14}.

Therefore, the proper procedure for constraining $\xi(m)$ is to pose
the hypothesis that there is a parent $\xi(m)$ from which the various
observed snapshots (e.g. the individual groups in Taurus-Auriga, or a
particular young or old star cluster) are drawn, thereby it being
essential to take into account in the analysis {\it all} biases and
evolution effects \cite{Kroupa13}. The mere counting-up of
observed ``stars'' (many of which are typically unresolved binary
systems) to create a histogram of masses, i.e. to obtain an estimate of the
stellar mass function, suggests such mass functions to have different
shapes. 
\begin{svgraybox}
  But careful analysis has always yielded the result that the
  hypothesis that there is one invariant parent distribution cannot,
  in most cases, be rejected, given all the uncertainties and biases.
\end{svgraybox}
\noindent This statement is true for star-formation that is and has
been occurring in the MW disk, including the Taurus-Auriga clouds and
most globular clusters, the Galactic field and bulge and dwarf
spheroidal satellite galaxies \cite{Kroupa13}.

Unless the job is done extremely carefully and thoughtfully, the
various outcomes of the star formation process will appear like a
mess, such that somewhat careless work may imply the result ``what you
see is what you get'', an {\it opinion} subscribed to by some workers.
But in this light the seminal~2007 paper by de~Marchi, Paresce \&
Pulone \cite{deMarchi07} reporting that low-concentration globular
clusters have present-day stellar mass functions which are depleted in
low-mass stars (i.e. they have bottom-light mass functions) came as a
shock. The dynamical clock ticks slower in low-concentration clusters,
such that the expectation was that these ought to, if anything, retain
the IMF at the low stellar-mass end. This is nicely shown by the
international collaboration led by Nathan Leigh (\cite{Leigh13}, their
fig.~4). This surprising observational result can be explained if
globular clusters were formed highly compact with radii smaller than
about~1~pc, more massive than today and with an IMF which
systematically becomes top-heavy with increasing birth density and
decreasing metallicity of the cluster with significant expansion
through the expulsion of residual gas \cite{Marks12}. The remarkable
finding by this study, led by Michael Marks, is that it is consistent
with the results obtained entirely independently from two studies led
by J\"org Dabringhausen concerning the dynamical $M/L$ ratios of and
the X-ray sources in ultra-compact dwarf galaxies (UCDs) \cite{Dab09,
  Dab10, Dab12}.  The dependency of the IMF on star-forming cloud
density and metallicity is shown in fig.~3 and 4 in \cite{Marks12}.
Furthermore, the first-ever integration of globular clusters on a
star-by-star basis over a full Hubble time by Akram Zonoozi et
al. furthermore significantly supports these results by uncovering the
initial conditions for the two clusters Pal~4 and Pal~14 after violent
revirialisation through gas expulsion \cite{Zonoozi11, Zonoozi14}.
The remaining challenge will be to see if the phase prior to violent
revirialisation is consistent with the above statements. The recent
constraints on the canonical shape of the low-mass stellar IMF in the
Arches star-burst cluster by Shin \& Kim \cite{ShinKim14} again
supports these results nicely. Further independent evidence for
top-heavy IMFs in extreme star-burst environments on scales of less
than 100~pc is seen in the high rate of type~II supernovae in
e.g. Abell~220 and~299 \cite{Perez09, Dab12, Kroupa13}.

Indeed, the concept of an invariant, universally valid parent IMF
stands in contradiction to all predictions star-formation theory has
been making over the past decades. According to even robust and
fundamental arguments in star formation theory, the IMF ought to
become top-heavy with decreasing metallicity and increasing gas
density and temperature.\footnote{\label{ftn_pk:bottomheavy}The recent
  much noted and important suggestion that the IMF becomes very bottom
  heavy with increasing mass of elliptical galaxies has been shown to
  be untenable (see \cite{SL13, Smith14, Peacock14}, with a possible
  solution to the spectroscopic evidence being proposed by
  \cite{Maccarone14}). Also, no theory of the IMF has ever {\it
    predicted} such a bottom-heavy IMF, while {\it predictions} were
  always such that the IMF becomes top-heavy under extreme conditions
  (e.g. Larson \cite{Larson98} as based on the Jeans-mass argument and
  Adams \& Fatuzzo \cite{AF96} as based on a self-regulated star
  formation theory).  No physical conditions are known which can
  generate such a bottom-heavy distribution of stellar masses
  (although \cite{Chabrier14} now suggest this may be possible, at
  least partially in highly turbulent high-Mach-number gas). Weidner
  et al. \cite{Weidner13c} point out the problems associated with such
  an IMF for the metal enrichment required to account for the observed
  abundances. Also, the relics of the most intense pc-scale star-burst
  systems known in the Local Group, the globular clusters, show
  bottom-light MFs \cite{deMarchi07, Leigh13} which can be accounted
  for only with significant dynamical evolution as noted above. The
  bottom-heavy IMF case will therefore not be discussed further here.}
Cases in point of theoretical IMF work investigating possible
variations with physical conditions are \cite{AF96, Larson98,
  Elmegreen99, Elmegreen00, PN02, Bate05, Klessen07, Bonnell07,
  Bate14, HB13}.

According to the above results gleaned largely from resolved stellar
populations, the following may be stated on the IMF:
\begin{svgraybox}
  The stellar IMF can be described as an invariant canonical
  distribution function (see also Sec.~\ref{sec_pk:IMFiswhat}) when
  the star-formation rate density (SFRD) in an embedded cluster is
  $\simless 0.1\,M_\odot/({\rm pc}^3\,{\rm yr})$, while it becomes
  progressively top-heavy with increasing SFRD \cite{Marks12}.
\end{svgraybox}

\subsection{IMF Question 2}
{\it Why it is so difficult to deduce the IMF of a galaxy}?

\subsubsection{Answer:}

Measuring the IMF of a simple resolved population is very challenging
(Sec.~\ref{sec_pk:eventIMF}), but deducing the IMF of a whole
star-forming galaxy is a very different problem. In a star cluster the
IMF can be constrained from the count of individual stellar systems
(single stars and unresolved binaries). For a galaxy this is not
possible, last not least because thee are far too many stars to count,
if stars can be resolved at all. Estimating the IMF of a whole galaxy,
the galaxy-wide IMF (GWIMF) or the IMF of a complex population, must
therefore rely on the integrated light properties of the galaxy, or on
spectroscopic analysis. {\it The former} can yield constraints on the
relative number of massive and less massive stars, since a population
with a top-heavy GWIMF will be blue, while a galaxy with a top-light
GWIMF will be redder. But there are degeneracies, such as younger
more-metal-rich populations being as red as old metal-poor populations
or populations with more bottom-heavy IMFs.  {\it The latter}
constrains the stellar population mixture more precisely from its
spectral energy distribution but relies on a template library of
stellar spectra which need to be combined in the correct proportions
to fit the observed SED. Ideally, all different methods would be used
in unison to enhance the constraints, but the workload is formidable
and subject to problems such as the spectral library not being
complete (if a type of star is not part of the library, other stars in
the library need to compensate its contribution which can bias the
result -- see footnote~\ref{ftn_pk:bottomheavy} on
p.~\pageref{ftn_pk:bottomheavy} for a possible example of this).
Also, in deriving the GWIMF it needs to be taken into account that
low-mass stars have been adding up over the star formation history of
a galaxy, while the massive star content is only visible as
established during a time corresponding to the life time of the
massive star being considered \cite{MS79, Scalo86,
  Kroupa13}. Normalisation issues between the low-mass end and the
high-mass end thus arise, as well as systematically different spatial
distributions between low-mass and massive stars. Low mass stars come
in ages extending to the birth of the galaxy and have thus had
many~Gyr to diffuse in phase space away from their original location
(e.g. the ancient thick disk), while those massive stars that were not
dynamically ejected from their birth clusters occupy the phase-space
region they were born in (e.g. the young thin disk). Similar issues
are dealt with in extreme detail by the seminal work on the IMF by
John Scalo \cite{Scalo86} and Bruce Elmegreen \& John Scalo
\cite{ES06}.

\subsubsection{The Galactic field IMF}
\label{sec_pk:galIMF}

The IMF of the MW disk can be constrained by carefully analyzing
direct star-counts. This is a difficult endeavor prone to biases
which, if not recognized, may affect the result to disadvantage. The
conversion of the stellar luminosity function to the stellar mass
function is proportional to the derivative of the stellar
mass--luminosity relation which has substantial uncertainties
\cite{Kroupa90}.  One can count the stars in dependence of their
absolute luminosity to construct the stellar luminosity function
within a small region around the Sun for which trigonometric parallax
is available. This ensemble of stars is so close by, within 5 to 20~pc
depending on the brightness of the star, that all multiple systems are
resolved such that an estimate of the individual stellar luminosity
function becomes possible. An alternative, in order to increase the
number of stars and thus the statistical significance of the stellar
count per luminosity bin, is to perform thin pencil beam surveys to
reach the stellar population along the line of sight out to~100 or
more~pc pioneered by Neill Reid and Gerry Gilmore. Many such pencil
beam surveys can be done, and distance measurements rely on the
photometric parallax method. Multiple systems remain unresolved.  The
biases associated with the two methods need to be understood very
well, and the structure of the Galactic disk needs to be modelled, as
well as the age and metallicity distribution of the stars of different
masses. Thus the Lutz-Kelker bias needs to be accounted for through
measurements errors in trigonometric parallax, cosmic scatter needs to
be modelled to account for Malmquist bias. The break-through seminal
paper on this problem has been contributed by Stobie, Ishida \&
Peacock in 1989 \cite{SIP89}.  A multi-dimensional minimisation
procedure, solving simultaneously for both types of star counts, has
been performed only once so far, in~1993 \cite{Kroupa93}.  The
resulting estimate of the IMF for main sequence stars with masses
below about $1\,M_\odot$ for the Galactic field population turned out
to be nicely consistent with Salpeter's work
\cite{Salpeter55}\footnote{Salpeter constrained the IMF for stellar
  masses in the range 0.4 to $10\,M_\odot$.}, and to be remarkably
robust over time and to be a good model for the parent IMF which is
consistent with the resolved stellar populations seen in current star
forming regions and in star clusters (see gray box on
p.~\pageref{p_pk:IMF}).  This result by \cite{Kroupa93} deviates from
the previous seminal work of Miller \& Scalo in 1979 \cite{MS79} and
Scalo in 1986 \cite{Scalo86} in that the mass--luminosity relation of
low mass stars was modelled physically properly for the first time
\cite{Kroupa90}, multiple systems were taken into account for the
first time \cite{Kroupa91}, and both, the nearby and the pencil-beam
surveys were combined consistently for the first and until now for the
last time \cite{Kroupa95MF}. The constraints of the field-star IMF by
\cite{Scalo86} remained valid for stars more massive than
$1\,M_\odot$, but this regime is very hard to treat because a
time-evolving star-formation history introduces structure into the
observationally derived IMF, as shown for the first time by Elmegreen
\& Scalo in 2006 \cite{ES06}.
\begin{svgraybox}
  Nevertheless, the overall slope of the field-star IMF above about
  $1\,M_\odot$, derived by Scalo's analysis \cite{Scalo86}, turned out
  to be steeper with $\alpha\approx 2.7$ \cite{Kroupa93} than the
  massive-star IMF deduced in individual very young populations,
  notably by the ground-breaking work of Phil Massey (see his review
  \cite{Massey03} and fig.~2 therein), $\alpha\approx2.3$,
  independently of metallicity and density for current star-forming
  regions \cite{Kroupa13}.
\end{svgraybox}
\noindent This difference between the field-star IMF and the IMF deduced in
star-forming regions remained unexplained for decades, and I simply
thought that the Scalo index may not be correct. In
Sec.~\ref{sec_pk:otherGal} this discrepancy will be resolved.

\subsubsection{Answer: the IMF in other galaxies}
\label{sec_pk:otherGal}

Deducing constraints on the GWIMF in external galaxies is hard because
one deals with integrated flux in various spectral pass bands, and
non-uniform extinction by dust, loss of photons, scattering of
photons, all play a role. Reducing the observations to a usable result
is a nightmare. But a few teams in the USA and in Australia have
managed break-throughs on this problem with rather dramatic results,
as will be touched upon further below in this section.

Observational evidence for a systematically top-heavy IMF in
star-bursting galaxies and regions therein and at larger redshift has
been suggested since decades (notably by Francesca Matteucci
\cite{Matteucci94}, see also \cite{Kroupa13} and references
therein). {\it But an underlying systematically varying and computable
  IMF model, which accounts for this observational evidence and at the
  same time also for the universality of the IMF in local star
  formation, was not available}. And, the observational evidence was
based on indirect arguments, such as the dynamical $M/L$ ratios of a
region, the available gas mass and its luminosity and the metallicity
distribution.  A computational approach did not exist at all, except
to make somewhat ad-hoc assumptions as to how the IMF may change with
redshift, for example, based on a Jeans-mass argument and ambient
temperature. The shape of the IMF remained unpredicted.

In any case, why should the IMF of a whole galaxy or of a large region
within it differ from the IMF in actual star-forming places which are
observed, wherever resolution is sufficient, to occur in pc-sized
cloud cores which may not know in which type of galaxy they condense
in out of a molecular cloud through self gravity?
\begin{svgraybox}
  One possible argument for a similarity between the IMF and the GWIMF
  would be if one assumes the IMF is a probability density
  distribution function. That is, in small pc-sized star-forming
  pockets $N_{\rm p}$ stars are drawn randomly from the same IMF as
  also describes the ramdom drawing process to form $N_{\rm g} \gg
  N_{\rm p}$ stars in a whole galaxy. Then, statistically,
  IMF$=$GWIMF. 
\end{svgraybox}
\noindent This (naive) ansatz was favored by most researchers,
including me (e.g. \cite{Elmegreen99, Elmegreen00, Kroupa01,
  Kroupa02b}, see also the discussion in \cite{Kroupa13}). But the
computational approach has changed dramatically through the discovery
of the IGIMF Theory in~2003 \cite{KW03}. The generic prediction of the
IGIMF theory that the GWIMF steepens at high stellar masses with
decreasing galaxy-wide SFR, has been confirmed by observations of
thousands of star forming galaxies \cite{HG08, Lee09, Meurer09,
  Gun11}.
\begin{svgraybox}
  As a result, neither the IMF nor the GWIMF are scale-invariant
  probability density distribution functions.
\end{svgraybox}
\noindent But before briefly explaining this computational approach it
is useful to address the perhaps most important observational evidence
which unambiguously indicates a systematic change of the GWIMF from
top-light at very low star formation rates (SFRs) to top-heavy at high
SFRs.  Surveys of hundreds and thousands of star-forming galaxies have
used various photometric tracers such as H$\alpha$ flux to test for
the high-mass end of the GWIMF, UV flux to test for the intermediate
mass stellar population and red broadband colors to test for the
intermediate and lower-mass end of the GWIMF \cite{HG08, Lee09,
  Meurer09, Gun11}. The data analysis and the investigations of
various biases such as from dust attenuation, loss of photons and
others, is highly involved and reported in these works in much detail.

\begin{figure*}
\centerline{
  \resizebox{14cm}{!}{
    \includegraphics[angle=-90,scale=1.0]{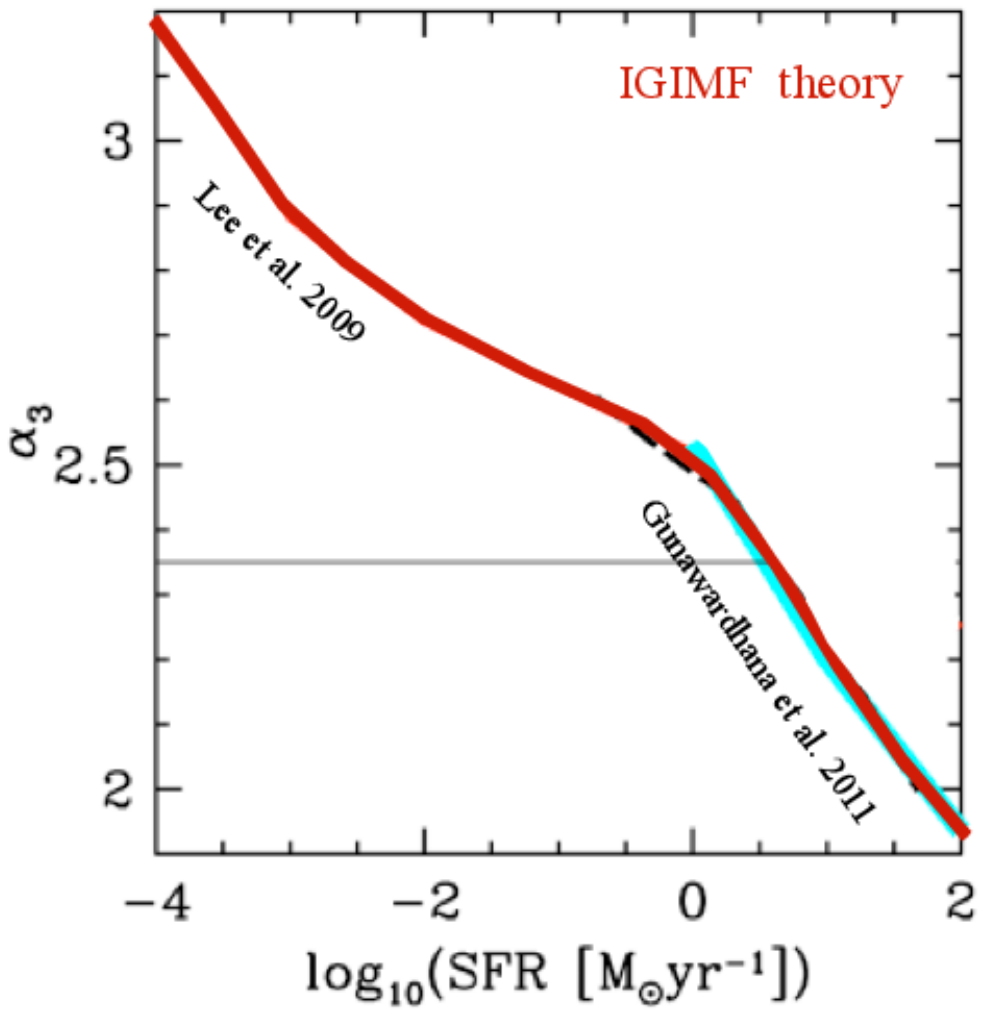}
 }}
\vspace{-10mm}
\caption{The power-law index $\alpha_3$ of the galaxy-wide IMF
  (GWIMF) for stars more massive than $\approx1\,M_\odot$ as a
  function of the galaxy-wide SFR is shown as the thick (red) solid
  line, as constrained by \cite{Lee09} for dwarf galaxies and by
  \cite{Gun11} for more massive galaxies, comparable and more massive
  than the MW. The solid curve coincides with the systematic variation
  of the IGIMF with SFR, as computed with the IGIMF Theory (adapted
  from fig.~1 in \cite{Weidner13b}. See also fig.~1 in Gargiulo et
  al. \cite{Gargiulo14}. The horizontal line marks the canonical
  Salpeter/Massey index $\alpha=2.35$.).
\label{fig_pk:IGIMF}}
\end{figure*}

The result in all of these surveys has been consistent in that the
GWIMF flattens progressively with increasing SFR. Modelling the GWIMF
as a canonical IMF which has $\alpha_1=1.3$ for stellar masses
$m<0.5,M_\odot$ and $\alpha_2=2.3$ for $0.5<m/M_\odot \simless 1$
with $\alpha_3$ being the index above $\approx 1\,M_\odot$, the
dependency of $\alpha_3$ on the SFR as deduced from the data is shown
in Fig.~\ref{fig_pk:IGIMF}.

How can this result of a systematically varying GWIMF with SFR be
understood in terms of the largely invariant stellar IMF deduced from
individual simple stellar populations (Sec.~\ref{sec_pk:eventIMF})?
\begin{svgraybox}
  The clue comes from realizing that the GWIMF is but the result of
  the addition of all simple populations in a galaxy to build-up the
  complex population of the galaxy. Thus, in simplified notation
  (SP$=$simple population $=$ embedded cluster $=$ star-formation
  event),
\begin{equation}
{\rm GWIMF}(m) = \Sigma_{{\rm SP}} \xi_i(m),
\label{eq_pk:IGIMF}
\end{equation}
where $\xi_i(m)$ is the stellar IMF contributed by the $i$th
star-formation event. 
\end{svgraybox}

How was this ansatz discovered? {\small In 2002 I was reconsidering my
  old problem (p.~\pageref{p_pk:hobby} above) of how thin galactic
  disks might thicken with time, and since I was working as a hobby on
  Nbody models of embedded star clusters which expel their unused gas
  through the action of their massive stellar content, I realized that
  such ``popping'' clusters may lead to hot kinematical components in
  the disk of a galaxy. Assuming all stars form in a distribution of
  embedded clusters, i.e. {\it that embedded clusters are the
    fundamental building blocks of galaxies} \cite{Kroupa05}, I
  calculated the integrals and found that it was readily possible to
  account for the thick disk and the subsequent thinning of the MW
  disk as time progressed if the SFR of the MW decreased with time
  until the present value of a few~$M_\odot/$yr \cite{Kroupa02}. This
  work done in 2002 constituted, without me knowing, the prediction that
  the MW would have been resembling a chain galaxy discovered in~2004
  by Bruce Elmegreen et al. \cite{Elmegreen04a}.}

With this ansatz, a similar integral over all star formation events or
all embedded clusters yielded the integrated galactic IMF
(IGIMF). Together with my then PhD student Carsten Weidner we did this
in~2003, finding that the Galactic-field IMF had to be steeper than
the IMF \cite{KW03}. This, of course, explained the result which Scalo
\cite{Scalo86} had already obtained (see Sec.~\ref{sec_pk:galIMF}).  A
generalization of this result to other galaxies became possible by
realizing that the most-massive cluster which is forming in a galaxy
depends on the SFR of the galaxy (\cite{Weidner04}, note the extension
to high SFRs by \cite{Rand13}).  This allowed us to make the
fundamental prediction that the IGIMF will flatten with increasing SFR
\cite{Weidner05}. This {\it predicted} behavior was confirmed {\it
  later} by the observational teams mentioned above. A particular
success was the prediction of the H$\alpha$ flux deficit over what is
expected for an invariant IMF for dwarf galaxies \cite{Pflamm09b}, as
confirmed by \cite{Lee09}.

One shortcoming of the IGIMF Theory as known then was that it could
not predict a top-heavy GWIMF, because the IGIMF could at most only
become as flat as the canonical stellar IMF (i.e. Salpeter index)
above $\approx 1\,M_\odot$ in the ``minimal scenario'' of Weidner \&
Kroupa \cite{Weidner05}. The knowledge of the top-heavy IMF in extreme
star-burst clusters discussed in Sec.~\ref{sec_pk:eventIMF} was not
known then. However, including that knowledge, which was obtained
entirely independently of the IGIMF Theory, into the RHS of
eq.~\ref{eq_pk:IGIMF}, yielded agreement with the observed GWIMF as a
function of SFR as shown in Fig.~\ref{fig_pk:IGIMF}. This was
published bei Weidner et al. in 2013 \cite{Weidner13b}. The
implication of this work is that the mass function of embedded
clusters (ECMF), i.e. of star formation events, needs to also become
somewhat top-heavy with an increasing galaxy-wide SFR. Galaxies with
high SFRs$> 10\,M_\odot$ thus also have slightly top-heavy ECMFs.

\subsection{IMF Question 3}
{\it Is the IMF really a universal function}?
\label{sec_pk:IMFiswhat}

\subsubsection{Answer: {\it Rules there are}}

This question can be answered readily today: {\it yes and no}: 
\begin{svgraybox}
  The IMF within a star formation event (i.e. embedded cluster) can be
  taken to be a mathematically defined parent distribution
  function, $\xi(m)$, which follows universal rules that make it
  dependent on the physical boundary conditions which determine the
  distribution function of star formation events that are physically
  accessible for a galaxy.
\end{svgraybox}
\noindent The parent distribution function of stellar masses formed in
one event (i.e. in an embedded cluster) is subject to conditions which
are axioms derived empirically (for a full list see
\cite{Weidner13b}):

\begin{svgraybox}
\begin{itemize}

\item For a star-formation rate density on a pc-scale SFRD$\simless
  0.1\,M_\odot/({\rm pc^3\, yr})$ the IMF is just the canonical form
  which can, for mathematical convenience, be written as a two-part
  power law form, or less conveniently as a log-normal part in the
  approximate range $0.08-1\,M_\odot$ (see gray box on
  p.~\pageref{p_pk:IMF}).

\item Based on independently obtained evidence from globular clusters
  and UCDs, the IMF becomes top-heavy when SFRD$\simgreat
  0.1\,M_\odot/({\rm yr}\;{\rm pc}^3)$ (Sec.~\ref{sec_pk:eventIMF}).

\item
The IMF is truncated at the canonical maximal stellar mass
$m_{\rm max*}\approx 150\,M_\odot$, as deduced by different
independently working groups (for the occurrence of {\it
  super-canonical stars} see \cite{Banerjee12}). 

\item The IMF, interpreted as an optimally sampled density
  distribution function \cite{Kroupa13}, has a most massive star which
  depends on the stellar mass of the star-formation event or embedded
  cluster, $m_{\rm max}={\cal K}_1(M_{\rm ecl})\le m_{\rm max*}$ (the
  $m_{\rm max}-M_{\rm ecl}$ relation). The function ${\cal K}_1(M_{\rm
    ecl})$ can be either fitted to the data or it may be derived
  independently from solving an integral equation (e.g. eq.~4-66 in
  \cite{Kroupa13}) and therefore directly follows from the shape of
  the IMF.

\item The most massive embedded star cluster forming in a galaxy
  depends on the SFR of the galaxy, $M_{\rm ecl,max}={\cal
    K}_2(SFR)$. Similarly to the function ${\cal K}_1$ above, this
  function ${\cal K}_2(SFR)$ can be fitted to data \cite{Weidner04,
    Rand13} or it may be derived independently from solving an
  integral equation which expresses the stellar mass forming in
  embedded clusters on the time-scale, $\delta t\approx 10\,$Myr,
  within which the inter stellar medium collapses to molecular clouds
  which then spawn the new population of stars (eq.~4.69 and~4.70 in
  \cite{Kroupa13}).

\item The mass function of star formation events or embedded clusters
  (the ECMF) becomes slightly top heavy when the galaxy-wide SFR$>
  1\,M_\odot/$yr (eq.~3 in \cite{Weidner13b}).

\end{itemize}
\end{svgraybox}

The {\it galaxy-wide IMF then follows from the above axioms} by
summing together all the IMFs contributed by each star formation event over
all star-formation events in a galaxy up to the most massive such
event which is sustained in the galaxy, given its SFR
(eq.~\ref{eq_pk:IGIMF}). This is the {\it IGIMF Theory}. It is a
theory because it is based on one principle, namely that star
formation always occurs in phase-space correlated star formation
events\footnote{These are the maxima in the density fluctuations in a
  turbulent molecular cloud, also called embedded clusters.  The
  least-massive examples of ``embedded clusters'' with a mass of about
  $5\,M_\odot$ are what some refer to as ``distributed star
  formation'', see the individual groups or clusters in Taurus-Auriga
  or in the southern part of the Orion L1641 cloud discussed above.}
and a small set of axioms derived from independent observations, and
because it is predictive. That is, with the IGIMF Theory it is
possible to calculate, from a few first principles deduced from
observation, how galaxies evolve, enrich with metals and buildup their
stellar masses (e.g. \cite{PAK08, Pflamm09b, Ploeckinger14, Recchi15,
  Gargiulo14}).

Self-similar \cite{Disney08, Speagle14, Tasca15} star-forming disk
galaxies are the by far dominant galaxy type \cite{Delgado10} above a
luminosity of $L\approx 10^{10}\,L_\odot$. The pronounced similarity
of galaxies is not expected in the SMoC (Standard Model of Cosmology)
\cite{Disney08} but is a manifestation of star formation being largely
self-regulated \cite{Koeppen95}, and this fundamental aspect of
galactic astrophysics is captured by the IGIMF Theory. The
top-heaviness of the IGIMF at very high SFRs (fig.~3 in
\cite{Weidner13b}, Fig.~\ref{fig_pk:IGIMF} above) immediately implies
that elliptical galaxies formed with top-heavy IMFs, in nice agreement
with the constraints on the IMF from the metal abundances brilliantly
deduced by Francesca Matteucci already in 1994
\cite{Matteucci94,GM97}).

\begin{svgraybox} {\it But why can the IMF not be a probability
    distribution function}?  Purely randomly sampling from a canonical
  IMF violates the too small spread in the IMF power-law indices
  deduced from many different simple populations by direct star counts
  (fig.~4-27 in \cite{Kroupa13}) and also the too small spread in the
  $m_{\rm max}-M_{\rm ecl}$ data, the spread in these data being
  consistent with measurement uncertainties \cite{Weidner13}. The
  physical spread thus seems to be small, such that the physical
  constraints required to ensure the small spread implies that even a
  probabilistically sampled IMF becomes indistinguishable from an
  optimally sampled IMF. The physical interpretation of this result is
  that star-formation appears to be highly self-regulated, in
  agreement with an attractive model of star formation by Adams \&
  Fatuzzo \cite{AF96}.  {\it Interpreting the IMF as an optimally
    sampled distribution function makes it mathematically convenient
    with the physical content of perfect self-regulation.}
\end{svgraybox}

Concerning the philosophical basis of the IGIMF Theory, there is a
nice little episode that occurred recently involving one of the
greatest minds in computational dynamics: in September 2014 I was
attending the workshop held in honor of Sverre Aarseth's 80th birthday
at an exclusive place in Sexten in the Dolomites. One day I was
walking with Seppo Mikkola and I mentioned to him ``{\it Nature must
  be surprisingly self-regulated}''. He replied unhesitatingly, ``{\it
  Yes, otherwise there would be complete chaos.}''  Indeed a direct
falsification of stochastic star formation has been achieved by an
investigation of the very young cluster distribution in the galaxy M33
by Pflamm-Altenburg et al. (2013, \cite{Pflamm13}).

\subsubsection{Stochastic community reactions}

Despite the rather impressive quality of the IGIMF Theory, it seems to
have implications which are unpalatable to parts of the community. One
is that the seminal Kennicutt relation \cite{Kennicutt94, Kennicutt98}
for calculating the SFR of a galaxy given its H$\alpha$ flux needs to
be corrected \cite{Pflamm07}. This centrally important relation for
extragalactic studies assumed the IMF to be invariant amongst
galaxies. But according to the IGIMF Theory, galaxies with a lower SFR
have a comparative deficit in their massive star content while the
Kennicutt relation was derived assuming an invariant ratio of massive
stars to low mass stars. This has deep implications for the
gas-depletion time scales and the stellar-mass buildup times of dwarf
galaxies \cite{Pflamm09}, which consequently do not fit the
present-day models tailored within the SMoC framework. With the IGIMF
Theory, a most remarkable prediction became possible, namely that
dwarf galaxies must have a smaller H$\alpha$/UV flux ratio than more
massive galaxies \cite{Pflamm09b}. The IGIMF Theory also predicts a
short radial cutoff of galactic disks in the H$\alpha$ flux, the disks
being much more extended in the UV \cite{PAK08}.  Both {\it
  predictions} are confirmed by observations \cite{Lee09, Boissier07}.

While we now have, for the first time, a computable IMF model which
encompasses universal star formation within the local smallest-groups
or ``distributed mode'' reaching up to major starbursts, it is amusing
but also frustrating to observe how parts of the community appear to
invest a very major effort to show that the IGIMF Theory is not
applicable. There is nothing to be written against critical tests. But
too many, and it seems all published work which claims to rule out the
IGIMF Theory I am aware of, has been shown to be flawed, either
because newer data made the original counter argument redundant, or
because the calculations are wrong.  It is worth considering these
reactions, since they imply that the community is now essentially
largely ignoring the IGIMF Theory for interpreting extragalactic
observations, rather than using the IGIMF Theory {\it as one possibility} to
interpret the observations.  For example, although \cite{Bruzzese14}
essentially find evidence for the IGIMF Theory by studying the stellar
population in the outer region of a dwarf galaxy, the IGIMF Theory is
not even mentioned, and instead stochastic star formation is used as
the favored model. This is done despite the evidence that stochastic
and unclustered star formation is not the appropriate description of
star formation in low-density regions (see \cite{Hartmann01, KB03,
  PS02, KM11}, fig.~1 in \cite{Hsu12}), and the explizit result that
stochastic star formation is ruled out given data \cite{Pflamm13,
  Kroupa13}.

A few cases in point which are fielded as arguments against the IGIMF Theory:

\begin{itemize}

\item In studying if a physical most-massive-star--star-cluster-mass
  ($m_{\rm max}-M_{\rm ecl}$) relation exists, \cite{MC08} write in
  their abstract ``Although we do not consider our compilation to be
  either complete or unbiased, we discuss the method by which such
  data should be statistically analysed. Our very provisional
  conclusion is that the data are not indicating any striking
  deviation from the expectations of random drawing.'' This one last
  sentence, {\it which only expresses an opinion}, does all the
  damage, as this paper is being cited as evidence against the
  existence of a physical $m_{\rm max}-M_{\rm ecl}$ relation. But
  \cite{MC08} culled their original data multiply times until they
  obtained a remnant distribution consistent with random selection of
  the most massive star from a model IMF, given $N$ stars in a
  model. That is, their modelling did not demonstrate that stochastic
  sampling from the IMF is a preferred model. Further, they did not
  test the hypothesis whether the $m_{\rm max}-M_{\rm ecl}$ relation
  is ruled out by their data, and their analysis is made redundant in
  any case by the new data obtained by Kirk \& Myers \cite{KM11} which
  show a very small spread at the low-mass end ruling out stochastic
  sampling \cite{Weidner13}.

\item Analyzing the spatial distribution of massive stars, \cite{PG07}
  argue that 4~per cent of O~stars which have been interpreted to have
  formed in isolation are consistent with stochastic/random sampling
  from the stellar IMF and therewith they argue against the existence
  of a physical $m_{\rm max}-M_{\rm ecl}$ relation. However, this
  exercise has become redundant because Gvaramadze et al.
  \cite{Gvaramadze12} have gathered data which show that virtually all
  of the previously thought 4~per cent ``isolated'' O~stars are most
  likely runaways. The remaining fraction of~O stars that cannot be
  identified as such is so small that it is not significant, but
  \cite{Gvaramadze12} demonstrate that it is consistent with the
  expected fraction of O~stars which cannot be traced back to their
  birth cluster due to the {\it two-step ejection mechanism}
  \cite{Pflamm10}. This mechanism operates by a massive binary being
  dynamically ejected from its birth cluster, and when the primary
  explodes as a supernova, the secondary is launched on a ramdom
  trajectory depending on the phase of its orbit. Thus agan, this
  ``evidence against a physical $m_{\rm max}-M_{\rm ecl}$ relation''
  does not stand up to scrutiny.

\item Notwithstanding the above rebuttals of the claims based on
  resolved populations fielded against the existence of a physical
  $m_{\rm max}-M_{\rm ecl}$ relation, \cite{Andrews13} deduce, from
  their observations of unresolved very young clusters in a distant
  dwarf galaxy, that the relation is not evident and that the IMF is
  randomly sampled. The problems their analysis suffers from are
  pointed out by \cite{Weidner14}, who show that once the analysis is
  done correctly, the same data in actuality are consistent with the
  physical $m_{\rm max}-M_{\rm ecl}$ relation. Not wavering in their
  quest to argue that the relation does not exist, they repeat their
  analysis in \cite{Andrews14} for another galaxy publishing a paper
  with significant text overlapt with the previous one.

\item There are other claims, none of which stand up to closer
  scrutiny, such as sometimes unwarranted criticism of the selection
  by Weidner et al of the $m_{\rm max}-M_{\rm ecl}$ data: the
  selection is based on two criteria only, namely the very young
  cluster has to be of age smaller than 4~Myr and must not have
  evidence for a supernova explosion, and the partially very large
  uncertainties are carried through properly into the analysis
  \cite{Weidner13}. Or, claims are put forward for cases of isolated
  massive star formation in nearby galaxies (such as in 2012,
  \cite{Bressert12}) as an argument for stochastic star formation
  based on oversimplified O-star propagation times, ignoring, for the
  sake of the argument it seems, that a major star-forming region
  contains many compact embedded clusters and that the two-step
  ejection mechanism pointed out in 2010 \cite{Pflamm10} leads to~O
  stars that cannot be traced back to their birth cluster. One of the
  authors of that study just said ``Who cares?'' when I pointed out
  that this mechanism most probably explains all their ``isolated''~O
  stars.

\end{itemize}

It is true that mistakes may happen, but these cases are mentioned
here as a documentation of possible evidence as to how the scientific
publications are sometimes designed in order to portray an opinion
rather than from evidence. Indeed, that the natural sciences have a
crisis is well known (see the gray box on p.~\pageref{pk:misconduct}
and Sec.~\ref{sec_pk:concs}), and the above suggests that astronomy is
not an exception.

Isolated massive star formation and the $m_{\rm max}-M_{\rm ecl}$
relation are central issues in the IGIMF Theory, because massive stars
can, according to this theory, only form in embedded clusters. It is
this relation which leads to galaxies with a low SFR, and which
therefore form low-mass embedded clusters only, to have a deficit of
massive stars compared to a statistically under-sampled IMF. Thus, if
it could have been shown that the $m_{\rm max}-M_{\rm ecl}$ relation
does not exist, then the IGIMF Theory with all its implications for
galactic astrophysics and cosmological star formation would not be
valid in the way it has been applied.\footnote{A weaker form of the
  IGIMF Theory persists nevertheless if it is assumed that all stars
  are formed in clusters which follow a cluster mass function. Only in
  the trivial and unphysical case that star formation is modelled as
  purely stochastic drawing of stars from an invariant IMF throughout
  a galaxy without further constraints would the IGIMF Theory imply
  IMF$=$IGIMF \cite{Weidner06} therewith violating the observational
  evidence that galaxies with a higher SFR have a systematically
  top-heavy IMF.} It is amusing to see how, as the evidence mounts
which demonstrates that the relation is physical, a few teams are
attempting to move their criticisms to ever more distant galaxies. For
example the attempts to prove that isolated massive star formation
does occur (which would violate the $m_{\rm max}-M_{\rm ecl}$ relation
here) in external galaxies where such opinions (rather than robust
calculations) can be barely disproven, given the extreme distances
involved, are heraldic. Or, publishing opinions in the abstracts of
peer-reviewed journal papers that unresolved very young (but partially
shrouded) clusters in distant galaxies disprove the existence of the
$m_{\rm max}-M_{\rm ecl}$ relation are comic at best.  Most
researchers do not have the time to analyse research papers in much
detail, and all too often the contents of an abstract are adopted
without careful perusal of the solidity of the contents. Thus opinions
may be propagated which lack a firm scientific foundation to, with
time, solidify a wrong but {\it majority view}.

At the end of the day, this situation is becoming as unsolvable as
someone claiming that Newton's law of universal gravitation is
falsified because in some distant apple trees there is evidence that
some apples did not actually drop down, thereby ignoring that unseen
animals devour the vanished apples. In this case the claim may not be
falsifiable if the animals are unobservable (too small, too quick).

\subsection{IMF Question 4}
{\it Does the IMF get heavier with galaxy mass, $M$, galaxy velocity
  dispersion,  $\sigma$, and metallicity $Z$}?

\subsubsection{Answer:}

Yes, it does. There is strong evidence suggesting that the IMF in
individual star formation events, i.e. in embedded clusters, becomes
top-heavy with increasing density and decreasing metallicity
(Sec.~\ref{sec_pk:eventIMF}; see footnote~\ref{ftn_pk:bottomheavy}
concerning the bottom-heavy IMF). The mathematical dependency on
density is stronger though, such that in extreme galaxy-wide star
bursts in which self-enrichment with metals from type~II supernovae
proceeds rapidly, the galaxy-wide IMF (GWIMF, eq.~\ref{eq_pk:IGIMF})
becomes top-heavy in galaxies with $SFR>1\,M_\odot/$yr. Massive
elliptical (E) galaxies are understood to have formed with very high
SFRs ($>10^3\,M_\odot/$yr) on a short ($<1\,$~Gyr) time scale, while
lower-mass E galaxies took longer to form \cite{Recchi09, Gargiulo14}.

\begin{figure*}
\centerline{
  \resizebox{9cm}{!}{
    \includegraphics[angle=0,scale=1.0]{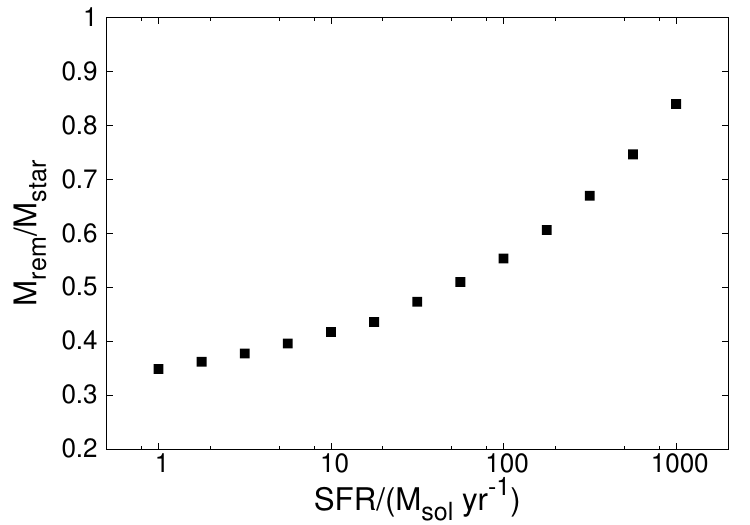}
 }}
\vspace{-0mm}
\caption{The fraction of mass in stellar remnants (white dwarfs,
  neutron stars, stellar black holes), $M_{\rm rem}$ divided by the
  total mass in shining stars with masses smaller than $0.8\,M_\odot$,
  as a function of the SFR of a galaxy. The IGIMF is calculated
  according to \cite{Weidner13b} assuming the mildly variable mass
  function of star formation events (i.e. of embedded clusters) and
  solar metallcity (see Fig.~\ref{fig_pk:IGIMF}). The production of
  stellar remnants is treated as in Dabringhausen et
  al. (\cite{Dab09}).  Kindly provided by Jan Pflamm-Altenburg.
  \label{fig_pk:Mrem}}
\end{figure*}

Thus, based on the IGIMF Theory it is expected that very massive
galaxies have a particularly heavy stellar population per unit light,
which consists of a substantial fraction of white dwarfs, neutron
stars and stellar mass black holes
\cite{Gargiulo14,Weidner13b}. Fig.~\ref{fig_pk:Mrem} shows the results
of an IGIMF model in which the metallicity is assumed to be solar.
Thus, a $10^{11}\,$yr old massive E galaxy weighing $10^{12}\,M_\odot$
in stellar mass which formed within 0.5~Gyr \cite{Recchi09,Gargiulo14}
would contain about as much mass in dark stellar remnants as in
shining stars, while a low-mass E galaxy ($10^8\,M_\odot$) which
formed with a SFR of $<1\,M_\odot/$yr would only have 35~per cent mass
in dark remnants in addition to its stellar mass. Detailed IGIMF
results on the dynamical $M/L$ ratios of E galaxies in comparison to
observational constraints are available in Gargiulo et
al. \cite{Gargiulo14}.  Because the mass, $M$, metallicity, $Z$ and
velocity dispersion, $\sigma$ of E galaxies are correlated positively
\cite{Cappellari12,Cappellari13}, a heavier IMF per unit light
correlates with larger $M, Z, \sigma$. Note that the previous result
reported by Cappellari \cite{Cappellari13} that more massive E
galaxies need a Salpeter IMF which has more faint (essentially dark)
M~dwarfs rather than a canonical IMF which has fewer M~dwarf stars is
degenerate with the alternative IGIMF Theory, namely a top-heavy GWIMF
with more dark remnants in more massive E galaxies and a GWIMF which
is closer to the canonical IMF for low-mass E galaxies.

A pioneering study in which the formation and evolution of E galaxies
in a SMoC Universe is studied self-consistently by employing the IGIMF
Theory has been made by Gargiulo et al. (2014)
\cite{Gargiulo14}. Their conclusions are rather remarkable, namely
that E galaxies appear to be better described by the
IGIMF theory rather than the customary invariant Salpeter IMF. They
emphasize in their Discussion ``In general, when the argument of a
variable IMF is considered to explain the [$\alpha$/Fe]-stellar mass
relation, the proposed IMF is treated as a free parameter . . . , or
is varied with exploratory aims . . . , following no particular
theory and leaving unexplored a vast region of the corresponding
parameter space. In this work, we test the well defined theory
regarding the integrated initial mass function of stars in galaxies
with top heavy IMFs in star clusters during starbursts''.

Concerning disk galaxies, the IGIMF Theory has been shown to reproduce
the observational constraints on how the GWIMF varies with SFR, as
discussed in Sec.~\ref{sec_pk:otherGal}.

\subsection{IMF Question 5}
{\it Why is the problem of the IMF related to the DM problem}?

\subsubsection{Answer:}

The IMF is related to the dark matter problem because a top-heavy IMF
yields dark stellar remnants which behave dynamically like cold dark
matter. Thus, as Fig.~\ref{fig_pk:Mrem} demonstrates, a massive E
galaxy which formed with a high SFR$> 1000\,M_\odot/$yr would contain
as much mass in dark stellar remnants as in shining stars.  An
astronomer analyzing the dynamical $M/L$ ratio assuming a universal
invariant IMF would wrongly conclude that the massive galaxy contains
dark matter. Further, the same hypothetical astronomer may also make
wrong deductions on the validity of Milgromian dynamics
(Sec.~\ref{sec_pk:Milgrom}).

How the galaxy-wide IMF affects fundamental physics is influenced
subtly in star-forming dwarf disk galaxies (i.e. dIrrs). These are
supposed to be dark matter dominated within their inner region. The
large cores of their putative dark matter halos are, however,
naturally and self-consistently explained in Milgromian dynamics
without dark matter \cite{FM12}. Now, in order to calculate the
contribution by stars to the potential, a galaxy-wide IMF is
required. If an invariant IMF is used for an ensemble of dIrr galaxies
which have different SFRs, the contribution by dark stellar remnants
would be calculated to be wrong, if instead the IGIMF Theory were the
correct description. Thus, a dIrr galaxy with an extremely low SFR
(say SFR$=10^{-4}\,M_\odot/$yr) would appear to have a redder stellar
population compared to a model with a canonical IMF, because the IGIMF
contains fewer massive stars at this SFR (Sec.~\ref{sec_pk:otherGal},
Fig.~\ref{fig_pk:IGIMF}). This may lead to errors in the age and/or
metallicity deduction, but will also affect the calculation of the
potential. If this is not taken into account, it may be concluded that
Milgromian dynamics does not work well in dIrr galaxies unless
Milgrom's constant $a_0$ is adjusted systematically with the mass of
dIrrs. This has indeed been found to be the case \cite{Randr14}, but
it is unclear at this stage whether using the IGIMF Theory would
alleviate this possible tension of Milgromian dynamics with the data.
Detailed modelling will be required to study this issue thoroughly.
\begin{svgraybox}
This highlights how the stellar IMF in galaxies affects
our ability of constraining fundamental physics.
\end{svgraybox}

\section{Conclusions}
\label{sec_pk:concs}
With this text two interrelated topics of galactic astrophysics,
namely the nature of gravitation and of stellar populations, have been
discussed critically. It has been argued that the most recent research
has unequivocally been showing that cold or warm dark matter particles
cannot be present. In the SMoC galaxies simply do not acquire the
observed properties. In reality galaxy--galaxy mergers are rare and
the motions of the satellite galaxies about the Milky Way cannot be
reproduced with dynamical friction on the expansive and massive dark
matter halos. In reality the vast number of galaxies are
rotationally-supported self-regulated systems in which baryonic
physics does not need to ``fight'' against the structures which
particle dark matter creates gravitationally. Feedback processes are
complex in detail, but readily computable.  Some of the galaxies
interact, an example being the possible Milky-Way--Andromeda encounter
about 11~Gyr ago which defined the present-day structure of the Local
Group. Only a few percent of the major galaxies are elliptical
galaxies which formed rapidly early on. Most dwarf satellite galaxies
appear to be tidal dwarf galaxies, the greatest fraction of which
formed early on. The empirical laws of galaxies are excellently
captured by two unifying principles, assuming the Standard Model of
Particle Physics (SMoPP) to be fully valid: the one is Milgrom's
scale-invariant dynamics in the weak field regime which is pointing
towards an advanced understanding of effective gravity, and the other
is the IGIMF Theory which follows from unifying principles derived
from well-observed local star formation processes. The following two
challenges may be useful for young researchers as a guideline for
deciding which type of research they would like to pursue:
\begin{svgraybox}
\begin{itemize}
\item Consider the observed distribution of light in an isolated disk galaxy:
  {\it predict} the rotation curve given only these data.
\item Consider a dwarf disk galaxy with a baryonic mass of $M_{\rm
    b}=10^8\,M_\odot$ and a massive disk galaxy with $M_{\rm
    b}=10^{11}\,M_\odot$ assuming they are isolated: {\it predict}
  their H$\alpha$ and their UV luminosities as well as their optical
  broad-band colors and magnitudes.
\end{itemize}
\end{svgraybox}

The reactions of parts of the cosmologically-relevant community (a) to
the evidence presented against the existence of particle dark matter
and (b) to the IGIMF Theory has been sketched. While it is crucial to
test any theory critically, these reactions are not entirely positive
nor productive, from a scientific point of view. Too much dogmatism
and peer-bullying appears to be evident, helping to keep younger
researchers away from following promising new but also somewhat risky
ideas; after all, we are trying to find out how the Universe
functions, and without trying all physical possibilities, how can we
know?  Particularly alarming is that even unphysical models are
propagated as major advances as long as these are encoded into the
SMoC framework.  Perhaps even more alarming are the documented poor
arguments fielded against, for example the relevance of phase-space
correlated satellite galaxy systems for testing the SMoC.  This hurts
scientific progress, last not least because funding for establishing
even small groups to study and test alternatives is hard to obtain.

A noteworthy schizophrene reaction of the scientific establishment to
new ideas seems to emerge: On the one hand side there is a generally
accepted model (the SMoC) which is based on a few empirically derived
axioms and which has a disastrous history of failed predictions. Here
parts of the community are putting an incredible amount of resources
in trying to fiddle some way out of the dis-functionality, whereby
none of the long list of failures on galactic scales and beyond have
until now ever been solved convincingly.  On the other hand side there
is a highly successful predictive theory based on a few empirically
derived axioms to calculate the properties of stellar populations.
But here parts of the community are putting an incredible amount of
energy in trying to argue that this theory is not applicable, probably
because long-lived culture would need to be revised.  Many of the
arguments invented for this purpose are weak, as shown by close
scrutiny, and do not serve to propagate an overly impressive quality
of astronomical research. Here, too, non-scientific methods appear to
be employed in attempting to suppress citations and use of the
relevant research literature.  

But why should the scientific establishment act in this way?  If I can
get a new insight into how nature functions I tend to be thrilled and
excited, even more so if my previous understandings or beliefs are
ruptured. But it seems that preserving the known order is also an
important driving force, and in some it may be more pronounced. The
organisation of funding systems plays a major role here: In the USA
the extreme competition for grants tied together with PI salary levels
discourages risky new ideas, while in Germany the strong hierarchy in
which virtually all institutional resources including secretarial are
the property of very few individuals for the time of their permanent
employment leaves little room for new groups to emerge following new
ideas. A discussion of directly related points on cosmology are
provided by Corrediora \cite{Corredoira09}, and The Economist has
recently rung alarm bells concerning the quality of research in the
Natural Sciences in general \cite{Fanelli10} and
\cite{TheEconomist13a,TheEconomist13b}, hereby not forgetting to
mention ``The trouble with physics'' raised by Lee Smolin
\cite{Smolin06}.

\begin{svgraybox}
  Astronomical and in particular cosmological research needs some
  improvement in attitude to count as a rigorous science. Opinions and
  beliefs appear to count too much. The notion that an abstract of a
  research paper is to portray an opinion of the authors seems to
  place such written record outside the realm of rigorous science.
  These statements appear to be relevant for all of the natural
  sciences, such that the notion that fundamental research is in a
  deep crisis cannot be negated with conviction.  This has
  implications for the long-term sustainment of our technological
  civilisation.
\end{svgraybox}

On the positive side, the above discourse on the failures of the SMoC,
the successes of Milgromian dynamics, the weaknesses of stochastic star
formation models, and the success of self-regulated star formation
including the IGIMF Theory, are opening great chances for young
researchers. There is an incredible opportunity to do fundamentally
important innovative research. The methods are available. For example, using the
Milgromian dynamics code ``Phantom of Ramses'' (POR) developed by
Fabian L\"ughausen in Bonn as a patch to RAMSES \cite{Lueghausen14} (see also
the similar release RAyMOND by Candlish et al. \cite{CSF15}), and
encoding into it the self-consistent star-formation recipes as
pioneered by Sylvia Pl\"ockinger in Vienna as a computational method of how
the IGIMF Theory self-consistently and naturally emerges in
star-formation simulations on galactic scales, it is now possible to
compute the formation and evolution of galaxies in an entirely
different cosmological framework and using much more realistic
baryonic processes than has ever been possible.\\

More specifically:  {\it What are the opportunities}? Here three are
mentioned (but please keep opinion out of the way): 
\begin{itemize}
  \item Test whether galaxies can form, be stable and look like the
    real things in a Milgromian dynamics universe without cold or warm
    dark matter. The simulation code POR has been developed for this
    purpose.

\item Develop a cosmological structure formation model which
  incorporates Milgromian dynamics and treats baryonic sub-grid
  processes self-consistently with the IGIMF Theory. A first step into
  this direction (but still in the Newtonian framework) have been
  taken by the work of Sylvia Pl\"ockinger et al. \cite{Ploeckinger14}
  and by Gargiulo et al. \cite{Gargiulo14}. 

\item Test the IGIMF Theory, by further investigating the $m_{\rm
    max}-M_{\rm ecl}$ relation and its physical contents, through
  direct counts of massive stars in nearby dwarf galaxies with very
  low SFRs, by applying it to study the SFR behavior of galaxies of
  all masses on a statistical basis, and by observing how the mass
  function of embedded clusters differs in galaxies with different
  SFRs and whether it is consistent with the IGIMF Theory.

\end{itemize}

It should be evident that the first person(s) providing full models of
galaxy formation and evolution in Milgromian dynamics are likely to
achieve significant break throughs, if these models turn out to
represent reality. After all, doing pure research is nothing else but
attempting to explore the unknown in order to understand and to
predict natural phenomena. It is risky, but also rewarding.

\bibliographystyle{unsrt}
\bibliography{/Users/pavel/PAPERS/BIBL_REFERENCES/kroupa_ref.bib}

\end{document}